\documentclass[reprint,prl,aps,superscriptaddress]{revtex4-2}
\usepackage{times}
\usepackage{graphicx}
\usepackage{amsmath, braket, amsfonts}
\usepackage{amssymb}
\usepackage{sidecap}
\usepackage{bm, color, ulem}
\usepackage{xcolor}
\usepackage{ragged2e}
\usepackage{wrapfig,lipsum,booktabs}
\usepackage{lineno}
\usepackage{comment}

\usepackage{siunitx}

\newcommand{\be}{\begin{equation}}
\newcommand{\ee}{\end{equation}}

\DeclareUnicodeCharacter{03BD}{{$\nu$}}

\makeatletter
\def\maketitle{
\@author@finish
\title@column\titleblock@produce
\suppressfloats[t]}
\makeatother

\begin{document}
\title{Nematicity and Orbital Depairing in Superconducting Bernal Bilayer Graphene with Strong Spin Orbit Coupling}
\author{Ludwig Holleis}  
\affiliation{Department of Physics, University of California at Santa Barbara, Santa Barbara CA 93106, USA}
\author{Caitlin L. Patterson}
 \affiliation{Department of Physics, University of California at Santa Barbara, Santa Barbara CA 93106, USA}
 \author{Yiran Zhang}   
 \affiliation{T. J. Watson Laboratory of Applied Physics, California Institute of Technology, 1200 East California Boulevard, Pasadena, California 91125, USA}
 \affiliation{Institute for Quantum Information and Matter, California Institute of Technology, Pasadena, California 91125, USA}
 \affiliation{Department of Physics, California Institute of Technology, Pasadena, California 91125, USA}
 \author{Yaar Vituri}
 \affiliation{Department of Condensed Matter Physics, Weizmann Institute of Science, Rehovot 76100, Israel}
 \author{Heun Mo Yoo}
 \affiliation{Department of Physics, University of California at Santa Barbara, Santa Barbara CA 93106, USA}
 \author{Haoxin Zhou}   
 \affiliation{Department of Physics, University of California at Santa Barbara, Santa Barbara CA 93106, USA}
 \email{Current address: Department of Electrical Engineering and Computer Sciences,
University of California, Berkeley, California 94720, USA.  }
 \affiliation{Institute for Quantum Information and Matter, California Institute of Technology, Pasadena, California 91125, USA}
 \author{Takashi Taniguchi}
 \affiliation{International Center for Materials Nanoarchitectonics,
 National Institute for Materials Science,  1-1 Namiki, Tsukuba 305-0044, Japan}
 \author{Kenji Watanabe}
 \affiliation{Research Center for Functional Materials,
 National Institute for Materials Science, 1-1 Namiki, Tsukuba 305-0044, Japan}
 \author{Erez Berg}
 \affiliation{Department of Condensed Matter Physics, Weizmann Institute of Science, Rehovot 76100, Israel}
 \author{Stevan Nadj-Perge}
 \affiliation{T. J. Watson Laboratory of Applied Physics, California Institute of Technology, 1200 East California Boulevard, Pasadena, California 91125, USA}
 \affiliation{Institute for Quantum Information and Matter, California Institute of Technology, Pasadena, California 91125, USA}
\author{Andrea F. Young}
\email{andrea@physics.ucsb.edu}
 \affiliation{Department of Physics, University of California at Santa Barbara, Santa Barbara CA 93106, USA} 
 
\date{\today}

\begin{abstract}

\end{abstract}

\maketitle


\textbf{
Superconductivity (SC) is a ubiquitous feature of graphite allotropes, having been observed in Bernal bilayers\cite{Zhou2022}, rhombohedral trilayers\cite{Zhou2021}, and a wide variety of angle-misaligned multilayers\cite{Cao2018,Park2021,Zhang2022b,Park2022}. 
Despite significant differences in the electronic structure across these systems, supporting the graphite layer on a WSe$_2$ substrate has been consistently observed to expand the range of SC in carrier density and temperature\cite{Arora2020,Zhang2023,Lin2022,Su2022}. 
Here, we report the observation of two distinct superconducting states (denoted SC$_1$ and SC$_2$) in Bernal bilayer graphene with strong proximity-induced Ising spin-orbit coupling. 
Quantum oscillations show that while the normal state of SC$_1$ is consistent with the single-particle band structure, SC$_2$ emerges from a nematic normal state with broken rotational symmetry. 
Both superconductors are robust to in-plane magnetic fields, violating the paramagnetic limit; however, neither reach fields expected for spin-valley locked Ising superconductors. We use our knowledge of the Fermi surface geometry of SC$_1$ to argue that superconductivity is limited by orbital depairing arising from the imperfect layer polarization of the electron wavefunctions. 
Finally, a comparative analysis of transport and thermodynamic compressibility measurements in SC$_2$ shows that the proximity to the observed isospin phase boundaries, observed in other rhombohedral graphene allotropes, is likely coincidental, constraining theories of unconventional superconducting pairing mechanisms in theses systems.}

Spin-orbit coupling (SOC) preserves the time reversal symmetry of electron bands in solids.  
As a result, SOC is not necessarily detrimental to the superconducting transition temperature: Cooper pairs may still condense from the degenerate Kramers' doublets by the same attractive interactions that lead to superconductivity in its absence\cite{Sigrist2009,Bauer2012}. 
Within a weak-coupling Bardeen-Cooper-Schrieffer picture, SOC may either raise or lower the density of states with opposite consequences for $T_c$. 
However, SOC does typically make superconductors more resilient to applied magnetic fields by pinning the spin direction of electrons.  
One example is Ising superconductivity\cite{Lu2015, Xi2016,Saito2016}, where in-plane mirror and time reversal symmetry protects Cooper pairs\cite{Mockli2020} making them, in theory, immune to arbitrarily large applied in-plane magnetic fields at zero temperature. Experimentally, however, other effects which break these symmetries will typically limit the critical in-plane magnetic field. 

Graphene\cite{Cao2018,Yankowitz2019,Hao2021,Park2021,Cao2021a,Zhang2022b,Park2022,Zhou2021, Zhou2022} provides a unique venue to investigate the interplay of superconductivity and spin-orbit coupling. Due to the small atomic number of carbon, the atomic SOC in graphene is small\cite{Sichau2019,Arp2023}.
However, SOC may be induced by supporting the graphene layers on a transition metal dichalcogenide substrate such as WSe$_2$\cite{Avsar2014,Wang2015,Wang2016,Yang2017,Volkl2017,Zihlmann2018,Wakamura2018,Wang2019}. 
The existing literature appears to show a systematic enhancement of superconducting transition temperatures for graphene systems with induced spin-orbit coupling.  
For example, twisted bilayer and trilayer graphene on WSe$_2$ substrates was observed to superconduct for a wider range of angles\cite{Arora2020, Lin2022}, while in twisted double bilayer graphene \cite{Burg2019,Shen2020,Najafabadi2020,Cao2020,He2021} superconductivity has been observed \textit{only} on WSe$_2$ substrates\cite{Su2022}.  
However, the lack of reproducibility in graphene moir\'e systems\cite{Lau2022} makes controlled experiments difficult.

 \begin{figure*}
    \centering
    \includegraphics[width = 150mm]{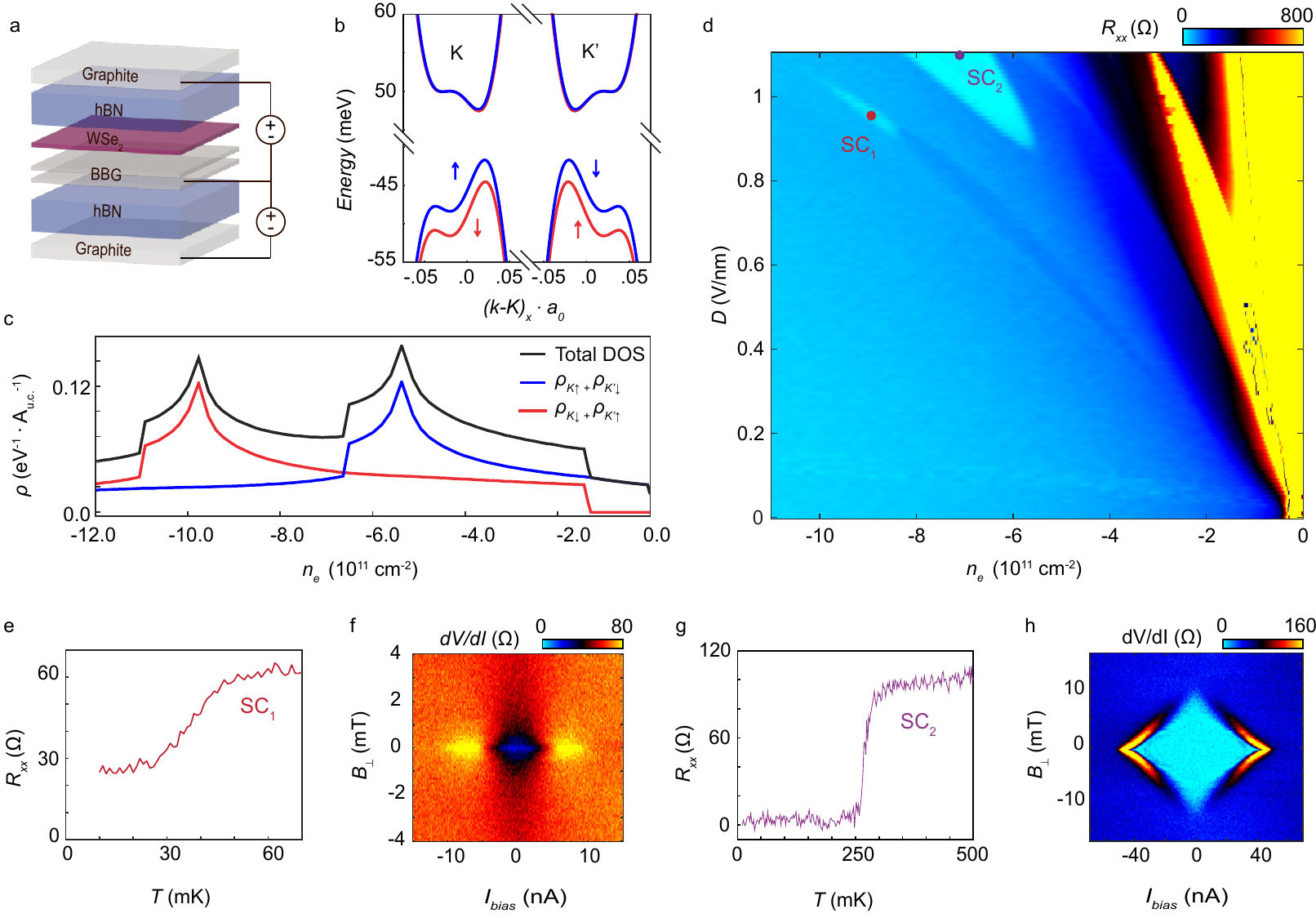}
    \caption{\textbf{Superconductivity in Bernal bilayer graphene (BBG) on WSe$_2$.} 
    \textbf{(A)} Sample schematic showing dual gated BBG on WSe$_2$. 
    \textbf{(B)} Band structure calculated within a tight binding model including Ising SOC. Bands correspond to the different isospin flavors as indicated.  Here $a_0$ = 2.46 \AA  is the graphene lattice constant. 
    \textbf{(C)} Total density of states (black) along with individual contributions from the spin/valley split bands, plotted for $D$ $\approx$ 1.0 V/nm.  
    \textbf{(D)} Longitudinal resistivity for hole-doped BBG/WSe$_2$.  Two superconducting phases SC$_1$ and SC$_2$ are marked. 
    \textbf{(E)} Temperature dependent resistivity of SC$_1$ at the point indicted by the red dot in panel D. 
    \textbf{(F)} Perpendicular magnetic field $B_\perp$ and current bias 
    $I_{bias}$ dependence at the same point.
    \textbf{(G)} Temperature dependent resistivity for SC$_2$ at the point indicted by the purple dot in panel D. 
    \textbf{(H)} $B_\perp$ and $I_{bias}$ dependence at the same point.}
    \label{fig:1}
\end{figure*}

Recently, it was shown that supporting Bernal bilayer graphene (BBG) on a WSe$_2$ substrate increases the maximal superconducting $T_c$ by an order of magnitude and dramatically expands the domain of carrier density and applied electric displacement field over which superconductivity is observed\cite{Zhang2023}. BBG is an ideal candidate to quantitatively study the effect of proximity induced SOC on superconductivity.   
First, the magnetic and superconducting phase diagram of  hexagonal boron nitride supported BBG is highly reproducible\cite{Zhou2022, Barrera2022}, allowing for reliable experimental controls for the effects of SOC.  
Second, the magnitude of the proximity-induced Ising SOC can be precisely determined \textit{in situ} using Landau level coincidences\cite{Island2019}.  
Finally, the simplicity of the BBG band structure allows for detailed comparisons between experiment and theoretical  calculations.  While prior experiments have found significant violations of the Pauli limit, the origin of the ultimate destruction of superconductivity in in-plane field has not been resolved, with both orbital effects and Rashba spin-orbit coupling possibly playing a role.

Here, we study a WSe$_2$-supported Bernal bilayer graphene device (Fig. 1A) with a measured proximity-induced Ising spin orbit coupling $\lambda_I=1.6$ meV (see Fig.~\ref{fig:S1}).  
We focus on hole filling and applied electric displacement fields $D>0$. 
In this regime, electronic states near the Fermi energy are polarized on the layer adjacent to the WSe$_2$\cite{Island2019,Zhang2023}.  
Fig.~\ref{fig:1}B shows the low energy band structure calculated within a tight binding model\cite{Jung2014} for an inter-layer potential of 100 meV, which corresponds to a displacement field $D$ $\approx$ 1 V/nm \cite{Zhang2009}. 
In the low density regime of $|n_e|$ $<$ 10$^{12}\mathrm{cm}^{-2}$, the measured Ising SOC is comparable to the Fermi energy, breaking the the native four-fold degeneracy of the spins and valleys and leaving a two-fold degeneracy between pairs of spin-valley locked bands. As shown in Fig.~\ref{fig:1}C, for $\lambda_I=1.6$~meV the single-particle density of states is characterized by two well-separated van-Hove singularities, corresponding to the saddle points in each degenerate pair of spin/valley locked bands.  Absent Ising SOC, the density of states of regular BBG displays only one van-Hove singularity\cite{Zhou2022}. 

\begin{figure*}
    \centering
    \includegraphics[width = 150mm]{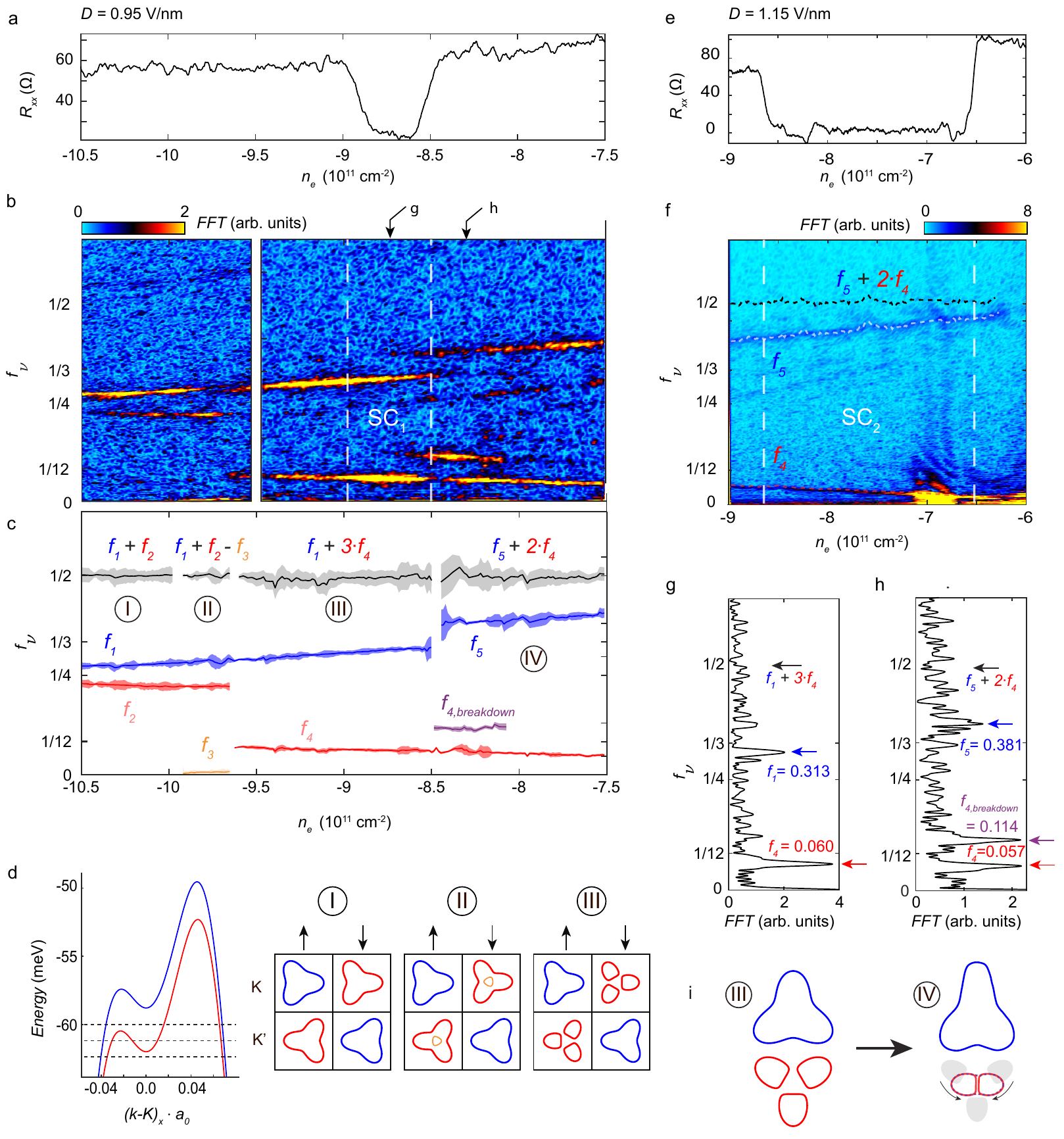}
    \caption{\textbf{Fermiology of the superconducting states in the presence of Ising SOC.}
    \textbf{(A)} R$_{xx}$ at $D$ = 0.95 V/nm, including the domain of SC$_1$. 
    \textbf{(B)} Fourier transform of $R_{xx}(1/B_\perp)$ over the same density range.  The Fourier transforms are performed over a field range of 130 - 400 mT and 130 - 260 mT in the left and right panels, respectively (see Fig.~\ref{fig:S3} and SI). 
    $f_\nu$ denotes the oscillation frequency normalized to the Luttinger volume, as described in the main text. 
    \textbf{(C)} Intensity peaks in $f_\nu$ from panel B. 
    Shaded areas represent error bars corresponding to the standard deviation of Gaussian fits to the peak frequencies. 
    The black lines show the indicated frequency sums, providing a visual representation of the sum rules according to Eq. \eqref{equ:1} described in the main text. 
    \textbf{(D)} Single-particle band structure near the $K$ point. The spin orbit split energy bands generate contrasting Fermi-surfaces for different spin-valley combinations, as shown in insets I, II, and III. 
    \textbf{(E)}  R$_{xx}$ at $D$ = 1.15 V/nm, including the domain of SC$_2$. 
    \textbf{(F)} $f_\nu$ corresponding to the $n_e$ and $D$ of panel E. 
    Dashed lines show the peak fits and the sum rule.  
    \textbf{(G)} Fourier transform amplitude of data in panel B for $n_e$ = -8.75 $\cdot$  10$^{11}$ cm$^{-2}$ 
    and 
    \textbf{(H)} $n_e$ = -8.3 $\cdot$  10$^{11}$ cm$^{-2}$.  
    The peak positions are indicated, illustrating the quantitative agreement of the contrasting sum rules.  
    \textbf{(I)} Schematic depiction of the nematic transition, in which one of the small Fermi pockets is absorbed by the large Fermi pocket with opposite spin and valley.}
    \label{fig:2}
\end{figure*}
 
Fig. \ref{fig:1}D shows electrical transport measurements for low hole densities and as a function of displacement field. 
For large displacement fields, we find two distinct superconducting states which we refer to as SC$_1$ and SC$_2$. 
SC$_1$ has a transition temperature $T_c\approx$ 40 mK, just above the base temperature of our dilution refrigerator.
As a result the resistance does not reach zero, showing a saturation at the lowest temperatures (Fig.~\ref{fig:1}E) that we attribute to disequilibration of the electron system with the phonon bath (see also Fig.~\ref{fig:S13}).  
However, nonlinear transport measurements in an applied perpendicular magnetic field (Fig.~\ref{fig:1}F) show both strong non-linearities at sub-10 nA currents and exceptional magnetic field sensitivity characteristic of low-$T_c$ superconductors in crystalline graphene systems\cite{Zhou2021,Zhou2022}.  
As shown in Fig.~\ref{fig:1}G, SC$_2$ has a much higher maximum transition temperature.  By fitting  the non-linear voltage to a Berezinskii–Kosterlitz–Thouless model \cite{Berezinskit1972, Kosterlit71973}, we find $T_{BKT}$ $\approx$ 255 mK (Fig.~\ref{fig:S2}).  The $B_\perp$ dependence shows a critical field of $B_C\approx 10 mT$ (Fig. \ref{fig:1}H), in line with previous reports of superconductivity in this regime\cite{Zhang2023}.  

\begin{figure*}[ht!]
    \centering
    \includegraphics[width = 150mm]{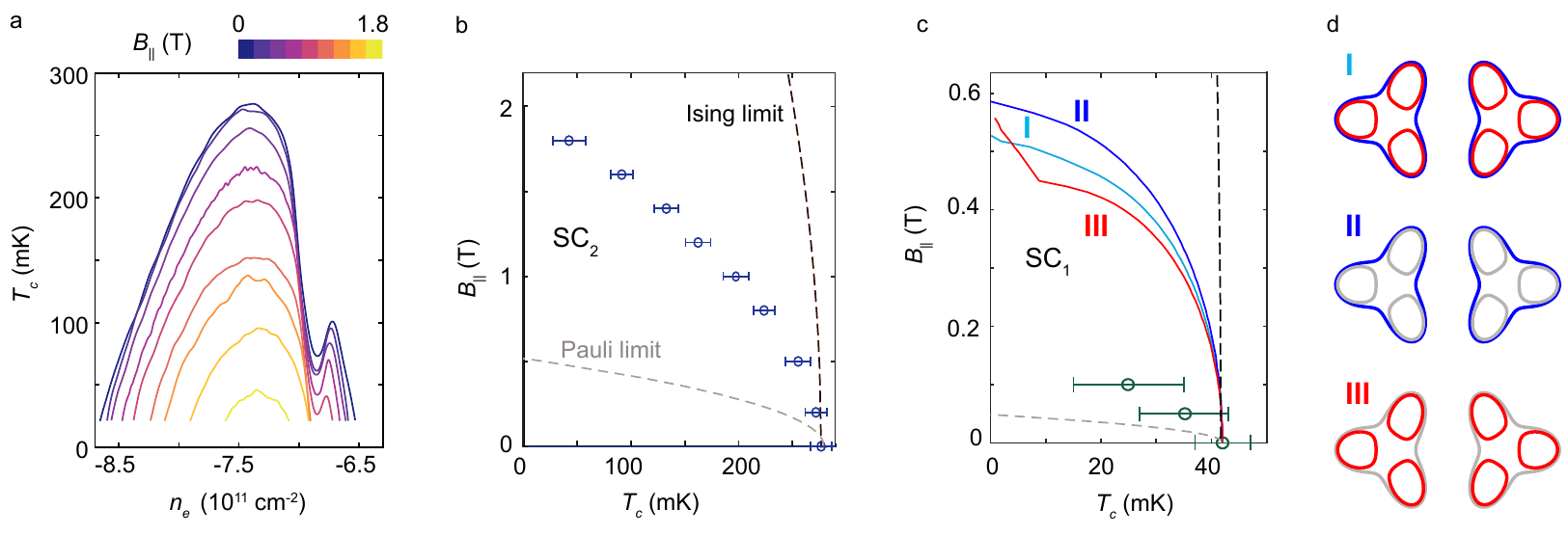}
    \caption{\textbf{Orbital depairing from in-plane critical fields:}
    \textbf{(A)} Superconducting transition temperature for different values of $B_\parallel$ for SC$_2$ at $D$ = 1.15 V/nm. 
    $T_c$ is extracted from temperature dependent resistance data (see Fig.~\ref{fig:S6}, \ref{fig:S7}) by defining $R(T_c)$ as half of the normal state resistance.  
    \textbf{(B)} $T_c$ as a function of $B_\parallel$ for SC$_2$ for the maximum of the T$_c$ domes in A. The black and grey dashed line represent the pure Ising superconductivity and Pauli limits, respectively.
    \textbf{(C)} The same plot for SC$_1$ where the bandstructure is in the approximate single particle limit known from quantum oscillations (see \ref{fig:2}). 
    The three solid lines represent the calculated in-plane critical fields including orbital depairing.
    Each line I-III represents a different subset of Fermi pockets hosting superconductivity as illustrated in D.
    \textbf{(D)} shows the single particle Fermi surfaces used to determine the in-plane orbital effect in C. 
    For I, II, III all Fermi pockets, only the large and only the three small pockets in each valley are included in the calculation, respectively. Fermi surfes not contributing to superconductivity are greyed out.}
    \label{fig:3}
\end{figure*}

To understand the normal phases from which these superconducting states condense, we perform high-resolution Shubnikov-de-Haas oscillation measurements, from which we resolve fine details of the Fermi surfaces.  
Fig.~\ref{fig:2}A, B show a comparison between zero magnetic field transport and the quantum oscillation frequencies $f_\nu$ at $D$ = 0.95 V/nm, across the domain of SC$_1$. Here, $f_\nu$ denotes the quantum oscillation frequency normalized to the total carrier density.
$f_\nu$ thus represents the fraction of the total Luttinger volume encircled by a given phase coherent orbit in momentum space. 
In these units, Luttinger's theorem may be expressed as 

\begin{align}
     \Sigma_{i} k_i \cdot f_{i}=1
    \label{equ:1}
\end{align}

where the index $i$ runs over all peaks corresponding to single orbits of a Fermi surface and $k_i$ is an integer degeneracy and carrier sign for the Fermi surfaces producing the oscillation at frequency $f_{i}$.  

Fig.~\ref{fig:2}C distills the data of Fig.~\ref{fig:2}B by plotting only the center frequencies and standard deviations  extracted from Gaussian fits of the observed peaks. 
Across the density range corresponding to regions I, II, and III, the oscillation frequencies can be understood from the SOC-modified single-particle band structure of Fig. \ref{fig:1}B-C. 
For example, region I shows two peaks with slightly different frequencies, denoted $f_1$ and $f_2$, where $f_1+f_2$ = 1/2 within experimental error. 
Using Eq.\ref{equ:1}, we see that Luttinger's theorem may be satisfied by assigning a two fold degeneracy to both Fermi surfaces, $k_1=k_2=2$.  As shown in Fig.~\ref{fig:2}D, this is consistent with SOC-modified band structure where the favored- and disfavored spin-valley locked bands have slightly mismatched Fermi surface areas even at comparatively high densities.  We denote this state Ising$_{2,2}$.
In region II, we observe the emergence of an additional peak, $f_3$. 
Due to its low frequency, $f_3$ is barely visible in Fig.~\ref{fig:2}B but can clearly be identified in the raw resistance and capacitance measurements (see Fig.~\ref{fig:S3}).
This is again consistent with band structure, and marks the formation of an annular Fermi sea with both hole-like and  electron-like Fermi surfaces. 
The sum rule implied by the band structure model, $f_1+f_2-f_3$ = 1/2 is again consistent with experimental data, and we denote this state Ising$_{2,2,-2}$. 
In region III, both $f_2$ and $f_3$ vanish and are replaced by a peak at intermediate frequency $f_4$. 
The Luttinger sum rule is satisfied for $k_1=2$ and $k_4=6$, as expected after the minority-occupation bands cross the saddle point and each annular Fermi sea breaks into three pockets.   
We label this state Ising$_{2,6}$, and conclude that SC$_1$ emerges from a normal state with no additional broken symmetries as compared to the single particle picture. 

At the low-$|n_e|$ boundary of SC$_1$, $f_1$ abruptly disappears and is replaced by a peak ($f_5$) at higher frequency and a peak at frequency $2\cdot f_4$, while $f_4$ continues its linear trend.  The fermiology of region IV is consistent across a large domain of $n_e$ and $D$, including the entire domain of SC$_2$, as shown in Figs. \ref{fig:2}E, F for the highest accessible displacement field and in Fig.~\ref{fig:S5} at $D$ = 1.05 V/nm for a larger density range. 
This is unexpected within the single particle picture, which cannot account for the emergence of a \textit{larger} $f_\nu$ peak as $|n_e|$ is lowered. 

As shown in Fig.~\ref{fig:S4}, the peak at $2\cdot f_4$ can be attributed to `magnetic breakdown'\cite{Biberacher2005} between the Fermi surfaces represented by $f_4$, and we denote it $f_{4,breakdown}$; as a result, it should not be counted towards the sum rule of Eq.\eqref{equ:1}. 
Remarkably, among simple possibilities, Luttinger's theorem is satisfied only for $k_5=2$ and $k_4=4$ (Fig.~\ref{fig:2}G, H). 
This assignment implies that for a single spin-valley flavor in a minority band, the Fermi surface degeneracy is 2---a fact plainly incompatible with preserving the C3 point group symmetry of bilayer graphene. 
We conclude that the parent state of SC$_2$, and adjoining state to SC$_1$, is nematic and we denote it N$_{2,4}$. 
Notably, most prior experiments probing possible nematicity in graphene devices\cite{Cao2021,Lin2023,Lin2022,Zhang2022c,Zhang2022d} have focused on resistance anisotropy. 
These measurements rely on structural uniformity of the device, a condition that can be difficult to meet under realistic experimental conditions in mesosopic two dimensional samples.
Because quantum oscillations probe  closed electron orbits in the sample bulk they provide a detection scheme for nematic order that is immune to many of these possible sources of systematic error.  Of course, the Fermi surface degeneracy is not  directly sensitive to nematic order, producing `false negatives' when its topology is compatible with $C_3$ symmetry. However, in the case studied here, quantum oscillations provide unambiguous evidence for a nematic ground state.  

Incidentally, attributing region IV to a nematic phase provides a natural explanation for the observation of the magnetic breakdown peak $f_{4,breakdown}$. Breaking $C_3$ symmetry relaxes the requirement that the small Fermi pockets be arranged symmetrically around the K and K' points. 
In this picture, the two small pockets remaining in the $N_{2,4}$ phase may move in momentum space to balance the competing effects of the kinetic and exchange contributions to the total energy; Fig.~\ref{fig:2}I illustrates a schematic representation of the fermiology in the $N_{2,4}$ phase near the III-IV transition where the pockets are very close, enabling breakdown. 
As the density is tuned deeper into region IV, the breakdown signal at $2\cdot f_{4}$ fades, implying that the pockets eventually decouple, growing farther apart in momentum space as they shrink in relative volume. \\ 

Theoretically, a variety of nematic phases have been proposed throughout the phase diagrams of Bernal and rhombohedral graphene\cite{Lemonik2010,Vafek2010,Mayorov2011,Dong2021,Huang2022,Szabo2022,Zhang2023, Curtis2022}. 
These include phases with differing isospin orders, including phases that conserve the occupation of the two valleys separately and those that develop inter-valley coherence\cite{Dong2021,Chatterjee2022,Huang2022,Zhang2023, Xie2023}.  
While quantum oscillations cannot distinguish these states directly, the phenomenology of SC$_2$, which develops in the $N_{2,4}$ phase, allows us to rule out at least some possibilities. \\

 Next, we study the in-plane critical field dependence of both superconductors. 
Fig.~\ref{fig:3}A shows $T_c$ as a function of the in-plane magnetic field, $B_\parallel$, at $D$ = 1.15 V/nm for SC$_2$. 
In Fig.~\ref{fig:3}B, we contrast the $B_{\parallel}$ dependence of the maximal experimental T$_c$ with two limits. The paramagnetic (``Pauli'') limit for a spin-singlet superconductor is determined by the competition between the Zeeman energy and the pairing gap (proportional to the transition temperature), $\mu_B B_P = 1.23 k_B T_{c,0}$\cite{Clogston1962,Chandrasekhar1962}.
For large Ising SOC (the ``Ising'' limit), in contrast, superconductivity is almost impervious to the in-plane Zeeman energy\cite{Lu2015} as long as the Zeeman energy is much smaller than the spin orbit coupling. 
In these systems, superconductivity consists of both spin singlet and spin-triplet components, with the latter becoming more dominant as the electron spins cant in an applied in-plane magnetic field.  

\begin{figure*}[ht]
    \centering
    \includegraphics[width = 150mm]{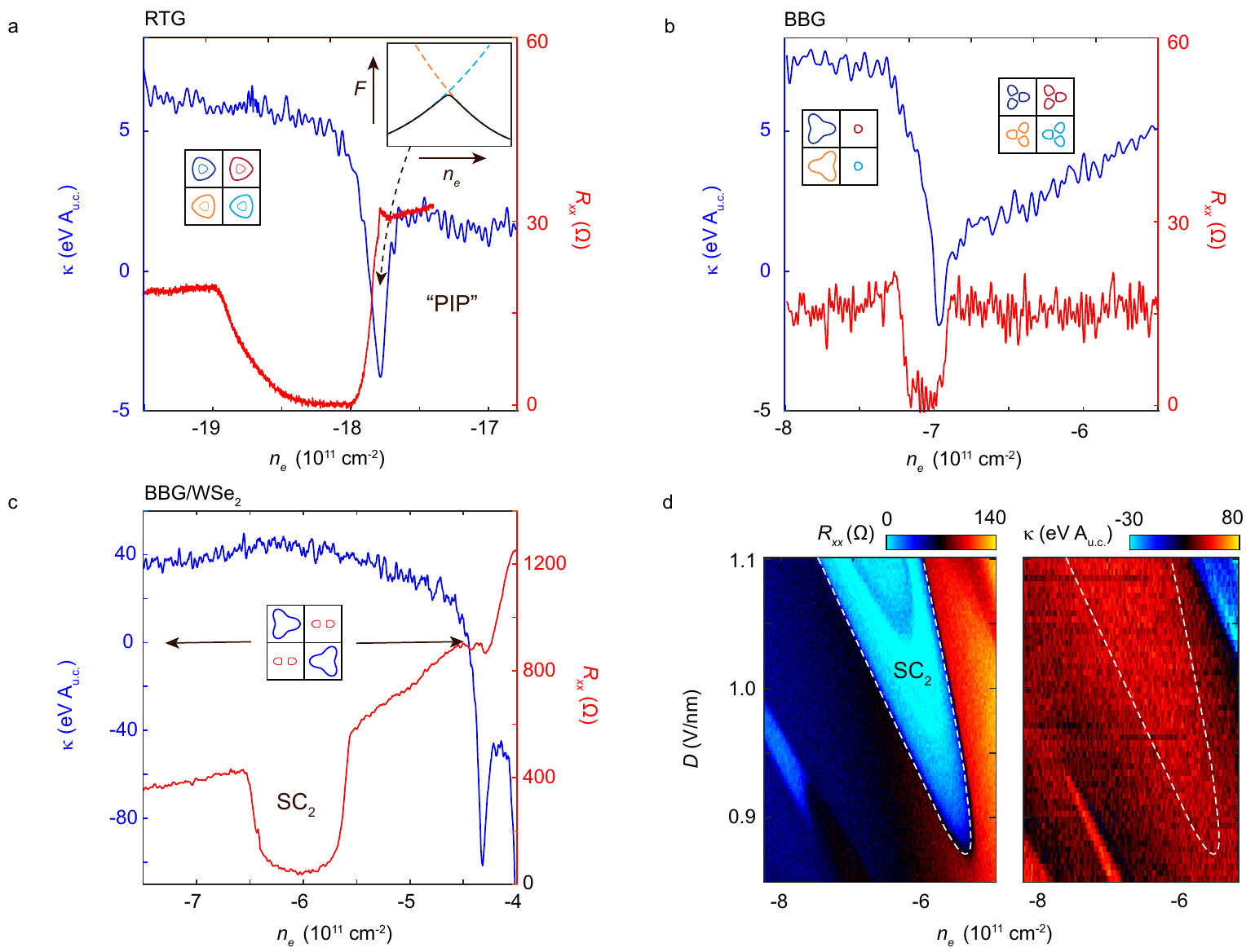}
    \caption{\textbf{Superconductivity and thermodynamic phase transitions in crystalline graphene superconductors}.
    \textbf{(A)} Resistivity $R_{xx}$ and inverse compressibility $\kappa=\partial \mu/\partial n_e$ in rhombohedral trilayer graphene (RTG) for $D$ = 0.46 V/nm and $T$ $\approx$ 60 mK\cite{Zhou2021}. 
    The lower inset illustrates the fermiology as described in \cite{Zhou2021}, while the upper inset depicts a schematic of the free energy (black line) across a first order phase transition.  As $\kappa=\partial^2 F/\partial n_e^2$, inverse compressibility is strongly negative at the transition. 
    \textbf{(B)} $R_{xx}$ and $\kappa$ for BBG, measured at $D$ = 1.04 V/nm and $B_{\parallel}$ = 400 mT.
    Insets depict fermiology as described in \cite{Zhou2022}. 
    \textbf{(C)}  $R_{xx}$ and $\kappa$ for for SC$_2$ in BBG/WSe$_2$ at $D$ = 0.95 V/nm taken at $T\approx 20$ mK. 
    The N$_{2,4}$ phase spans nearly the entire range of the plot.  
    \textbf{(D)} $R_{xx}$ (left) and $\kappa$ (right) measured over a range of $n_e$ and $D$. 
    The superconducting dome is overlaid from the transport onto the inverse compressibility map via the white dashed line.}
    \label{fig:4}
\end{figure*}

Our experimental data fall between these two limits.  This raises the question of which of several possible mechanisms--including Rashba spin orbit coupling, disorder, and the orbital effect of the in-plane field\cite{Mockli2020, Saito2016, Zhang2023}--limit superconductivity at high $B_\parallel$. 
To address this question, we focus on SC$_1$, which also shows a $B_\parallel$ dependent critical temperature intermediate between the Pauli and Ising limits (see Fig. \ref{fig:3}C).  
In contrast to other superconducting states observed in Bernal bilayer graphene, however, the fermiology of the normal state 
of SC$_1$ extracted from quantum oscillation measurements is well fit by a single particle band structure model (See Fig. \ref{fig:S19}).  This implies that  Coulomb interactions induce only moderate Fermi surface renormalization in this regime.  We may thus take a single particle band structure model as a basis to estimate pair breaking effects.  To model the pair-breaking effect of the in-plane magnetic field, we set $\vec{A}=z(\vec{B}\times \hat{z})$ and perform a Peierls substitution to extract the magnetic field dependent band structure. 
We take a linear approximation of the dispersion in $\vec{B}$ and in the momentum perpendicular to the Fermi surface and solve the linearized gap equation to find $T_c$ as a function of $B_\parallel$. 
As shown in Figure \ref{fig:3}C, including the orbital effect of $B_\parallel$ lowers the critical magnetic field from $B\approx 10T$ at low temperatures to $B_\parallel \approx 500 mT$, accounting for the bulk of the discrepancy with the experimental data.  
Rashba spin-orbit coupling induced by the WSe$_2$ substrate may also be included within this model; for Rashba coupling constant as large as $\lambda=\SI{2}{meV}$, we find that it has negligible effect on $T_C$ due to the sublattice polarization of the electronic states near the Fermi level (see fig. \ref{fig:S20}). 
Notably, although the dispersion and the orbital moment are not symmetric to in-plane rotations, we find numerically that $T_c$ is only weakly dependent on the direction of $B_\parallel$ in the plane, consistent with a lack of detectable experimental dependence on the angle between $B_\parallel$ and the graphene lattice vectors. 
This leaves the question of the origin of the remaining quantitative discrepancy between the experimental data and our model.  One source of systematic error is experimental: the low values of $T_C$ will be affected by the disequilibrium between the charge carriers with the phonon bath, which tends to decrease measured $T_C$ relative to its true value.  

An additional discrepancy may arise from the fact that quantum oscillations measure only the Fermi surface pocket size and number, but not their position. Coulomb-induced Fermi surface reconstructions may change the positions of the Fermi pockets--and thus the in-plane orbital  moment of states at the Fermi surface--without noticeable changes to fermiology inferred from quantum oscillations. To investigate the possible magnitude of this effect, we compare three models for superconducting pairing (see Fig. \ref{fig:3}C-D) based on the Fermi surfaces calculated from the single-particle band structure. These assume  superconducting pairing within all, only the large, or only the small Fermi pockets.  These different assumptions produce a range of  low temperature critical $B_\parallel$ of $\sim 150mT$, despite the close proximity in momentum space of the different Fermi pockets.  It seems likely, then, that additional Fermi surface reconstruction may be sufficient to close the gap between theory and experiment.

We conclude that orbital effects are the dominant source of depairing in in-plane magnetic fields.  Notably, this effect applies to both the Ising-enhanced and spin-triplet superconductors in graphene: the contrasting in plane orbital magnetic moments in the two valleys provide a source of pair breaking in a magnetic field for any superconductor in which pairing occurs between states at the two corners of the Brillouin zone.  This finding implies that detailed knowledge of the Fermi surface is essential for interpreting the in-plane critical field in multi-layer graphene superconductors--besides violations of the Pauli limit, the upper critical field may not provide any information on the spin structure of the condensate. Rather, Pauli limit violations may provide information about the momentum space positions of the Fermi surfaces involved in superconducting pairing.  For example, both rhombohedral trilayer and BBG/WSe$_2$ devices\cite{Zhou2021,Zhang2023} have shown Pauli limit violations that become largest before the system undergoes a phase transition; this may imply that pairing between states is occurring closer to the Brillouin zone corners, where the wave functions are most layer polarized.

Lastly, we contrast superconductivity in BBG/WSe$_2$ with crystalline graphene superconductors without proximity induced SOC. In these systems, superconductivity has been observed only near isospin phase transitions\cite{Zhou2021,Zhou2022}. However, the nature of these transitions has not been experimentally determined. To characterize the phase transitions and their connection to superconductivity, we measure both resistance and the inverse electronic compressibility $\kappa = \partial \mu / \partial n_e$.  We take pains to measure these simultaneously, ensuring that compressibility measurements are taken under identical conditions, and do not heat the electron systems (see Fig. \ref{fig:S14} and Methods).  
Fig.~\ref{fig:4}A shows the comparison of $R_{xx}$ and $\kappa$ for rhombohedral trilayer (RTG). The isospsin transition between the isospin unpolarized state and the partially isospin polarized state (PIP) is first order, as indicated by the negative compressibility peak (see Fig. 4A inset) observed at the phase boundary.
In this regime, superconductivity is observed on the isospin-disordered side of the phase boundary. 
Fig.~\ref{fig:4}B shows the same comparison for BBG, taken in the high-D, high $B_\parallel$ regime where triplet superconductivity is observed. 
Again, superconductivity occurs immediately adjacent to a first order phase transition, in this case on the isospin-ordered side. 

In BBG/WSe$_2$, the evolution of the isospin polarization near SC$_1$ bears a superficial resemblance to these systems, with superconductivity developing adjacent to a phase boundary (Fig.~\ref{fig:S12}). 
However, the phenomenology of SC$_2$ contrasts markedly: as illustrated in Fig.~\ref{fig:4}C, superconductivity emerges deep within the nematic N$_{2,4}$ phase, far from any isospin phase transitions. 
This is corroborated by Fig.~\ref{fig:4}D, which compares transport and compressibility across a wide range of $n_e$ and $D$; the domain of superconductivity approaches a first order isospin transition only at the largest displacement fields, but is otherwise uncorrelated.
The lack of direct correlation between superconductivity and phase boundaries is also evident in the  quantum oscillation data of Fig.~\ref{fig:2}F, where superconductivity develops, peaks, and subsides over a range for which fermiology evolves monotonically. 

These observations suggest that the observed correlation between superconductivity and phase transitions in crystalline graphene layers is likely coincidental. 
Rather, superconductivity evolves within a given phase but may be destroyed if a competing phase with different symmetry becomes energetically favorable. 
Within this picture, the enhancement of superconductivity in WSe$_2$-supported graphene devices arises from the broken `spinless' time reversal symmetry relating wave-functions in opposite valleys with the same spin. 
Breaking this symmetry at the single-particle level preferentially stabilizes a subset of orders featuring two-particle states near the Fermi level compatible with Cooper pairing\cite{Sigrist2009,Bauer2012}. It then becomes possible to trace the evolution of superconductivity within a single phase to its maximum strength. 
SOC may also suppress fluctuations of electronic orders which may suppress superconductivity\cite{Curtis2022,Xie2023,Wagner2023, Chou2022a, Jimeno-Pozo2022,Shavit2023}.  
In the case of the valley-symmetric nematic order depicted in Fig.~\ref{fig:2}I, $n_e$- and $D$-tuned details of the electron wavefunctions near the Fermi surface must then account for non-monotonic dependence of the transition temperature across the phase.   

By breaking a symmetry not required for superconductivity (spinless time reversal), SOC stabilizes electronic orders with higher superconducting $T_c$, including  nematic states.
In the future, this procedure may be generalized, for example by applying uniaxial strain.  Coupling of the strain to a nematic order parameter would then increase the range of stability of the $N_{2,4}$ phase, possibly further enhancing superconductivity.

\textbf{Acknowledgements.} 
The authors would like to acknowledge  discussions  with Maksym Serbyn, Areg Ghazaryan, A. H. MacDonald, Zhiyu Dong, Maxim Khodas, and Patrick A. Lee. 
The work was supported by the Office of Naval Research under award N00014-20-1-2609, and the Gordon and Betty Moore Foundation under award GBMF9471. 
Work at Caltech has been supported by the NSF-CAREER award (DMR-1753306).  
K.W. and T.T. acknowledge support from the Elemental Strategy Initiative conducted by the MEXT, Japan (Grant Number JPMXP0112101001) and JSPSKAKENHI (Grant Numbers 19H05790, 20H00354 and 21H05233). E.B. and Y.V. were supported by NSF-BSF award DMR-2310312 and by the European Research Council (ERC) under grant HQMAT (grant agreement No. 817799).

\bibliographystyle{apsrev4-1}
\bibliography{references}


\clearpage
\newpage
\pagebreak

\onecolumngrid

\begin{center}
\textbf{\large Supplementary information }\\[5pt]
\end{center}

\setcounter{equation}{0}
\setcounter{figure}{0}
\setcounter{table}{0}
\setcounter{page}{1}
\setcounter{section}{0}
\makeatletter
\renewcommand{\theequation}{S\arabic{equation}}
\renewcommand{\thefigure}{S\arabic{figure}}
\renewcommand{\thepage}{\arabic{page}}

\textbf{This PDF file includes:}
\begin{itemize}
    \item Materials and Methods
    \item Band Structure Calculations
    \item Additional Notes on the Competing Order State
    \item Supplementary Figures
\end{itemize}

\section{Materials and Methods}

\textbf{Sample preparation:} the bilayer graphene, monolayer WSe$_2$ (commercial source, HQ graphene) and hBN flakes for the van-der-Waals heterostructure are obtained by standard mechanical exfoliation of bulk crystals. 
The heterostructure is assembled via a dry transfer technique\cite{Wang2013} using poly bisphenol a carbonate (PC) placed on a polydimethylsiloxane (PDMS) stamp. 
A dual graphite gated device design is chosen to reduce the charge disorder \cite{Zibrov2017} and tune density $n_e$ and displacement $D$ field independently. 
The geometry of the device is then defined by a CHF$_3$/O$_2$ etch and contacted by ohmic edge contacts of Ti/Au (5nm/100nm). 
The BBG/WSe$_2$ sample and other devices of the same geometry were also studied in \cite{Zhang2023} for more information on sample preparation. 
For details on sample fabrication of the rhombohedral trilayer and Bernal bilayer graphene without WSe$_2$ measured in Fig.~\ref{fig:4}, see Ref. \cite{Zhou2021, Zhou2021a, Zhou2022}. 
Sample 2 of Ref. \cite{Zhou2022} is used here to study BBG without WSe$_2$ support.

\textbf{Transport measurements:} 
Longitudinal resistance measurements are performed utilizing lock-in techniques at frequencies in the DC limit $<$ 48 Hz and low currents of 1 - 2 nA. 
Each transport line is filtered by several Pi- and RC-filter stages at the mixing chamber and an additional RC filter with a cutoff $<$ 5.7 kHz on the sample holder to reduce electronic noise and lower the electron temperature. 
We want to note that the present device has very long ($>$ 10 $\mu$m) dual gated bilayer graphene contact leads (sample image see \ref{fig:S13}A) - compared to the many layer graphite contacts of devices used in previous studies\cite{Zhou2022}. 
Thus, the contacts are in the same phase as the bulk of the sample which has two consequences: first, contact resistances are a higher than for many layer graphite contacts of similar aspect ratio, leading to additional heating at the contacts. 
We attribute the saturation of resistance of SC$_1$ at the lowest temperatures to this heating effect (see Fig.~\ref{fig:1}D) while the  disequilibration of the charge carriers with the phonon bath simultaneously prevents efficient cooling. 
Second, the contacts become superconducting at much lower critical currents (corresponding to similar current \textit{densities}) due to their long aspect ratio. 
In Fig.~\ref{fig:S13}B, raw data is shown taken at $n_e$ = -7.3 $\cdot$ 10$^{11}$ cm$^{-2}$, $D$ = 1.15 V/nm where $T_c$ of SC$_2$ is maximal. 
The AC voltage drop across the sample and the DC current flowing through the sample is plotted against applied DC voltage bias.
While the raw voltage drop follows typical behavior for a superconductor, the DC current shows two kinks without corresponding feature in the AC voltage data. 
These kinks can be explained by the superconducting transition of the dual gated bilayer graphene leads.
As we calculate $R_{xx} = V_{AC}/I_{DC}$, such kinks visually imprint themselves onto the noise fluctuations of the zero resistance state---even in a four-terminal measurement setup.
Therefore, features within the superconducting state such as in e.g. Fig.~\ref{fig:2}E or Fig.~\ref{fig:4}D should be disregarded as artefacts.

\textbf{Shubnikov-de-Haas measurements:} 
Magnetic field dependence of the longitudinal resistance of Fig.~\ref{fig:2}, \ref{fig:S3} and \ref{fig:S5} are taken with constant spacing in 1/$B_\perp$ down to the lowest fields where quantum oscillation are visible. 
We perform Fourier transformations over different field ranges (see \ref{fig:S4}) sampled in $1/B_\perp$ and show the lowest field range in the main text for which all primary peaks of the quantum oscillations are clearly visible. 
This method reduces effects of higher harmonics and magnetic breakdown between different Fermi surfaces\cite{Biberacher2005} (additional peaks in Fig. S\ref{fig:S4}C) and ensures that we can correlate our QO data with the zero field phase diagram.  To convert gate voltages into charge density and normalize frequencies to the total Luttinger volume (as done in Fig.~\ref{fig:2}, \ref{fig:S3} and \ref{fig:S5}), we calibrate the geometric capacitance of the gates to the sample via Shubnikov-de-Haas oscillations at zero displacement field and large densities. Further, a displacement field dependent voltage offset has to be accounted for. This voltage offset arises from that fact that the chemical potential has to be shifted out of the band gap of the BBG before increasing gate voltages is going to increase the charge carrier density. As transport and penetration field capacitance measurements are inaccurate in determining the exact position of the band edge in gate voltage space due to large resistances and long charging-times, we instead utilize the fully spin- and valley-polarized phase. Its frequency $f_\nu$ should be exactly one by definition, as all charge carriers live on a single Fermi surface. We adjust the voltage offset so that this is true over the whole gate voltage range of the quarter metal. 

\textbf{Field dependence:} 
All measurements are performed in a dilution refrigerator with base temperature of $\sim$ 10 - 20 mK and a 1T/1T/9T (XYZ) superconducting vector magnet. For the measurements of $T_c$ domes vs $B_{\parallel}$, the sample is mounted with the field of the Z magnet aligned to the bilayer graphene plane and the out-of plane field component is carefully minimized to be $<$ 0.1 mT using the XY magnets. 
$T_c$ is determined for discrete in-plane fields as flux trapping in the XY magnets and field misalignment of the Z magnet with the sample in this configuration prohibits continuous sweeping of $B_{\parallel}$ while keeping $B_\perp$ $\approx$ 0.0 mT.
Numerical values of $T_c$ in Fig. \ref{fig:3} are extracted from fits of a step function to the raw data in fig. \ref{fig:S6}, \ref{fig:S7}, \ref{fig:S8}. We define $T_c$ as the temperature when the resistance is half of its normal state value.

\textbf{Penetration field capacitance:} 
We utilize a capacitance bridge circuit to measure penetration field capacitance. 
In a previous capacitance bridge setup (\cite{Zibrov2017}), a FHX35X high electron mobility transistor (HEMT) is directly connected to the sample gate, resulting in elevated electron temperatures above 100 mK. 
In order to lower the electronic temperature, we add an isolation capacitor made of silicon between the HEMT and the sample. 
We also installed a series of attenuators at each temperature stage to ensure thermalization of input coaxial cables. 
Finally, we added an ATF34143 HEMT at the 4K stage to transform the output impedance of the FHX35X HEMT and increase the bandwidth of output signal.  (for full circuit schematic see Fig. ref{fig:S14}).

We apply an A.C. excitation of $\sim$ 1 - 2 mV and a frequency of 166.667 kHz. 
The improved capacitance setup described above allows us to measure inverse compressibility at the base temperature of our dilution unit. 
This can be best seen in Fig.~\ref{fig:4}B where transport and capacitance are measured simultaneously. 
Superconductivity is well developed in this case. This indicates no significant increase in the electron temperature; as a reference,  $T_c$ in BBG on hBN is less than 30mK. 
Indeed, $R_{xx}$ linecuts with and without\cite{Zhou2022} simultaneous capacitance measurement show no noticeable difference.

\section{Band Structure Calculations}
We computed the electronic dispersion $\epsilon_{\vec k,j}$ and density of states $\rho(u,n_e)$ of BBG with layer-specific Ising spin-orbit coupling at interlayer potential $u = -100$ meV and electronic density $n_e$. 
The band structure of bilayer graphene without spin-orbit coupling is well-described near the $K$ and $K'$ points by a continuum model, expressed in the $(A1,B1,A2,B2)$ basis as \cite{McCann2006,Jung2014}

\begin{align*}
   \mathcal H_0 = \begin{pmatrix}
    -u/2 & v_0 \pi^\dagger & -  v_4 \pi^\dagger &  v_3 \pi  \\
    v_0\pi & -u/2 &\gamma_1 & - v_4 \pi^\dagger \\ -v_4 \pi & \gamma_1 & u/2 & v_0 \pi^\dagger\\ v_3\pi^\dagger & -v_4 \pi & v_0\pi & u/2
    \end{pmatrix},& &\pi = \hbar (\xi k_x + i k_y),\quad v_i = \frac{\sqrt{3}a}{2\hbar}\gamma_i.
\end{align*}

Here, $a = 2.46$ \r{A} is the monolayer graphene lattice constant and $\xi  = \pm 1$ indicates whether $\mathcal H_0$ has been expanded about the $K=\left(\frac{4\pi}{3a},0\right)$ or $K'=\left(-\frac{4\pi}{3a},0\right)$ valley. 
We extract the band structure parameters from fits to quantum oscillation data (fig. \ref{fig:S19}).
The hopping parameters are: $\gamma_0$ = 2880 eV for same layer hopping between nearest-neighbor sites, interlayer coupling term $\gamma_1 = 361$ meV for same effective hopping between dimer sites $B_1$ and $A_2$, interlayer coupling term $\gamma_3 = 323$ meV for trigonal warping, and $\gamma_4 = 30$ meV for interlayer coupling between the dimer and non-dimer sites. Adding monolayer WSe$_2$ adjacent to one side of the bilayer graphene induces Ising SOC on the adjacent layer and can be accounted for in the model by adding a Hamiltonian $\mathcal H_I$ \cite{Khoo2018}

\begin{align*}  
    \mathcal H_I = \begin{pmatrix}\xi \lambda_I \hat s_z  & 0& 0 & 0\\
    0 & \xi\lambda_I \hat s_z & 0 & 0 \\ 0 & 0 & 0 & 0 \\ 0 & 0  & 0 & 0\end{pmatrix}
\end{align*}

acting on the $A_1$ and $B_1$ sublattices. 
Here, $\hat s_z$ is a Pauli matrix acting on the spin subspace. 
The total Hamiltonian is then $\mathcal H  = \mathcal H_0 + \mathcal H_I$.
The electron density $n_e(\mu, u)$ at chemical potential $\mu$ and interlayer bias $u$ is computed by evaluating 

\begin{align*}
    n_e(\mu, u) =  \left(\frac{dk}{2\pi}\right)^2\sum_{\vec k,j}\frac{1}{\exp\big(\beta (\epsilon_{\vec k,j} - \mu)\big) + 1}
\end{align*}

where $j$ indexes the energy bands and $\vec k$ runs over the regular grid of momenta on which we computed the electronic dispersion $\epsilon_{\vec{k},j}$.
The factor $dk^2$ normalizes for grid spacing. 
The electron density in Fig.~\ref{fig:1} was evaluated at temperature $T$ = 10 mK.
The density of states $\rho(\mu, u)$ is computed by grouping the energy eigenvalues in bins of equal width $dE$ and normalizing by bin size to be consistent with $dn_e/d\mu$.

\section{$T_c$ calculations including orbital magnetic field}

To include the orbital effect of in-plane magnetic field in the single-particle dispersion, we take $\vec{A}=z(\vec{B}\times \hat{z})$ and perform a Peierls substitution, which amounts to taking 

\begin{align*}
   \mathcal H_0 (\vec{B})= \begin{pmatrix}
    -u/2 & v_0 \pi_+^\dagger & -  v_4 \pi^\dagger &  v_3 \pi  \\
    v_0\pi_+ & -u/2 &\gamma_1 & - v_4 \pi^\dagger \\ -v_4 \pi & \gamma_1 & u/2 & v_0 \pi_-^\dagger\\ v_3\pi^\dagger & -v_4 \pi & v_0\pi_- & u/2
    \end{pmatrix},& \quad 
    \begin{matrix}
        \pi_+ = \xi (\hbar k_x+\frac{edB_y}{2}) + i(\hbar k_y-\frac{edB_x}{2}), & v_i = \frac{\sqrt{3}a}{2\hbar}\gamma_i \\
        \pi_- = \xi (\hbar k_x-\frac{edB_y}{2}) + i(\hbar k_y+\frac{edB_x}{2}), & \pi = \hbar (\xi k_x + i k_y),
    \end{matrix} 
\end{align*}
with $d$ being the interlayer distance.

For simplicity, we consider a singlet pairing interaction local in the continuum (valley) description
\begin{equation*}
    \mathcal{H}_\text{int} = -V_s \int d^2 r [\psi_+^\dagger(-is^y)\psi_-^\dagger][\psi_-(is_y)\psi_+],
\end{equation*}
where $\psi_\pm$ is the 8-dimensional (spin, layer, and sublattice) vector of annihilation operators of a particle in valley $\xi=\pm1$, and $s_y$ is the second Pauli matrix in spin space. The gap equation is given by

\begin{equation*}
    \Delta_{\vec{q}} = \frac{V_s}{\Omega}\sum_{\vec{k}}\langle \psi_{\vec{k}+\vec{q}/2,-}(is^y)\psi_{-\vec{k}+\vec{q}/2,+} \rangle_{\Delta_{\vec{q}}}, 
\end{equation*}
where $\Omega$ is the system's area, and the expectation value is computed with respect to the mean field Hamiltonian $\mathcal{H}_{\text{MF}}$ given below.

Denoting the transformation to band basis as 
\begin{equation*}
    \psi_{\vec{k},\alpha,s,\xi}=\sum_n u^{\alpha,s}_{n,\vec{k},\xi}c_{n,\vec{k},\xi}, \qquad \alpha=\{\sigma,l\}\; \text{(sublattice and layer index)},
\end{equation*}
and neglecting inter-band pairing (justified due to the small SC gap at the vicinity of the transition), the mean-field Hamiltonian is given by

\begin{equation*}
    \mathcal{H}_\text{MF}=\sum_{n,\vec{k},\xi}\varepsilon_{n\vec{k}\xi} c_{n\vec{k}\xi}^\dagger c_{n\vec{k}\xi} -\Delta_{\vec{q}} \sum_{n,\vec{k},\alpha,s,s'} (u^{\alpha,s}_{n,\vec{k}+\vec{q}/2,+})^*(-is^y)_{s,s'}(u^{\alpha,s'}_{n,-\vec{k}+\vec{q}/2,-})^* c_{n,\vec{k}+\vec{q}/2,+}^\dagger c_{n,-\vec{k}+\vec{q}/2,-}^\dagger+\text{h.c.},
\end{equation*}
where 
\begin{equation*}
    \Delta_{\vec{q}} = \frac{V_s}{\Omega}\sum_{n,\vec{k},\alpha,s} (u^{\alpha,s}_{n,-\vec{k}+\vec{q}/2,-})(is^y)_{s,s'}(u^{\alpha,s'}_{n,\vec{k}+\vec{q}/2,+})\langle c_{n,-\vec{k}+\vec{q}/2,-}c_{n,\vec{k}+\vec{q}/2,+} \rangle, 
\end{equation*}
or in its linearized form

\begin{equation*}
    1 = \frac{V_s}{\Omega}\sum_{n,\vec{k}}|f_{n,\vec{k},\vec{q}}|^2\frac{ \tanh(\frac{\beta \varepsilon_{n,\vec{k}+\vec{q}/2,+}}{2}) +\tanh(\frac{\beta \varepsilon_{n,-\vec{k}+\vec{q}/2,+}}{2})}{2(\varepsilon_{n,\vec{k}+\vec{q}/2,+}+\varepsilon_{n,-\vec{k}+\vec{q}/2,+})}, \qquad f_{n,\vec{k},\vec{q}}=\sum_{\alpha,s,s'}(u^{\alpha,s}_{n,\vec{k}+\vec{q}/2,+})^*(-is^y)_{s,s'}(u^{\alpha,s'}_{n,-\vec{k}+\vec{q}/2,-})^*.
\end{equation*}

It is important to note that the pairing potentials $\Delta_{\vec{q}}$ at different momenta $\vec{q}$ are decoupled in the linearized gap equation. Thus, one can calculate $T_c^{(\vec{q})}$ for each momentum $\vec{q}$  independently and take $T_c=\max_{\vec{q}} (T_c^{(\vec{q})})$. For finite in-plane magnetic field, time-reversal symmetry is broken, and therefore one really has to scan $\vec{q}$ and cannot simply assume $T_c^{(\vec{q})}$ to be maximal for $\vec{q}=0$. Generically, as one can expect, we find the optimal $\vec{q}$ to point along the direction of $\vec{B}\times \hat{z}$.

In practice, to achieve results with high enough momentum resolution at a reasonable run time, we expand the non-interacting spectrum as
\begin{equation*}
    \varepsilon_{n,\vec{k}_F+\delta \vec{k},\xi}(\vec{B})=\vec{v}_{n,\xi}(\vec{k}_F)\cdot \delta \vec{k} +\vec{\mu}_{n,\xi}(\vec{k}_F)\cdot \vec{B},
\end{equation*}
where $\vec{k}_F$ is a point on the Fermi surface of the $n^{th}$ band at $\vec{B}=0$, $\delta \vec{k}$ is a vector pointing in the perpendicular direction from the Fermi surface, $\vec{v}_n(\vec{k}_F)=\vec{\nabla}_{\vec{k}}\varepsilon_{n,\vec{k}_F,\xi}$ is the Fermi velocity and $\vec{\mu}_{n,\xi}(\vec{k}_F)=\vec{\nabla}_{\vec{B}} \epsilon_{n,\vec{k}_F,\xi}$ is the orbital magnetization of the Bloch state at a point on the Fermi surface labeled by $\vec{k}_F$.

\section{Additional Notes on the Competing Order State}

In the main text discussion of Fig.~\ref{fig:3}, we have noted a suppression of superconductivity at densities around -6.9 $\cdot$ 10$^{11}$cm$^{-2}$ and highest displacement fields. 
Here, additional data is shown in order to support and contextualize this argument and give further information about this "competing order state" (COS). 
First, the location of the COS is shown in Fig.~\ref{fig:S10} --- only developing at high $D$ above 1.05 V/nm for a small density range. 
A small out-of-plane magnetic field favors such state over SC$_2$ as indicated by the black arrows.
Fig.~\ref{fig:S9}A-D displays temperature and $B_\perp$ dependence of SC$_2$ at $D$ = 1.15 V/nm. 
Again, the COS appears at finite out-of-plane field and elevated temperatures with a characteristically increased resistance. 
It is useful to compare the energy scales of superconducting gap $\Delta$ and depairing energy $E_D$ due to finite Cooper pair momentum.
For a simple BCS superconductor, the superconducting gap is proportional to $k_B T_c$.
An out-of-plane magnetic field breaks Cooper pairs due to their orbital motion at an energy that is proportional to $\sqrt{H_{c,2}}$.
Thus, we contrast $T_c^2$ and $H_{c,2}$ plotted against density.
Their density dependence agrees reasonably well within error, most notably in the region of the COS at $n_e$ $\approx$ -6.9 $\cdot$ 10$^{11}$cm$^{-2}$, see fig. \ref{fig:S9}C.
Following this argument, not only $T_c$, but the superconducting gap is suppressed at these densities.

Non-linear transport reveals that COS is destroyed at finite current (Fig.~\ref{fig:S9}E-G) with a critical temperature similar to $T_c$ of SC$_2$, reminiscent of a charge density wave state.
Interestingly, the balance between COS and SC$_2$ is tipped in favor of superconductivity by lowering the magnetic field and temperature (Fig.~\ref{fig:S11}).
We also want to note that the fermiology evolves smoothly across COS (Fig.~\ref{fig:2}F) implying no additional symmetry breaking within the different isospin flavors. 
Instead, a natural explanation might involve nesting of different Fermi surfaces---only favored for wave vectors at specific densities.

\clearpage

\section{Supplementary Figures}

\begin{figure*}[h]
    \centering
    \includegraphics[width = 150mm]{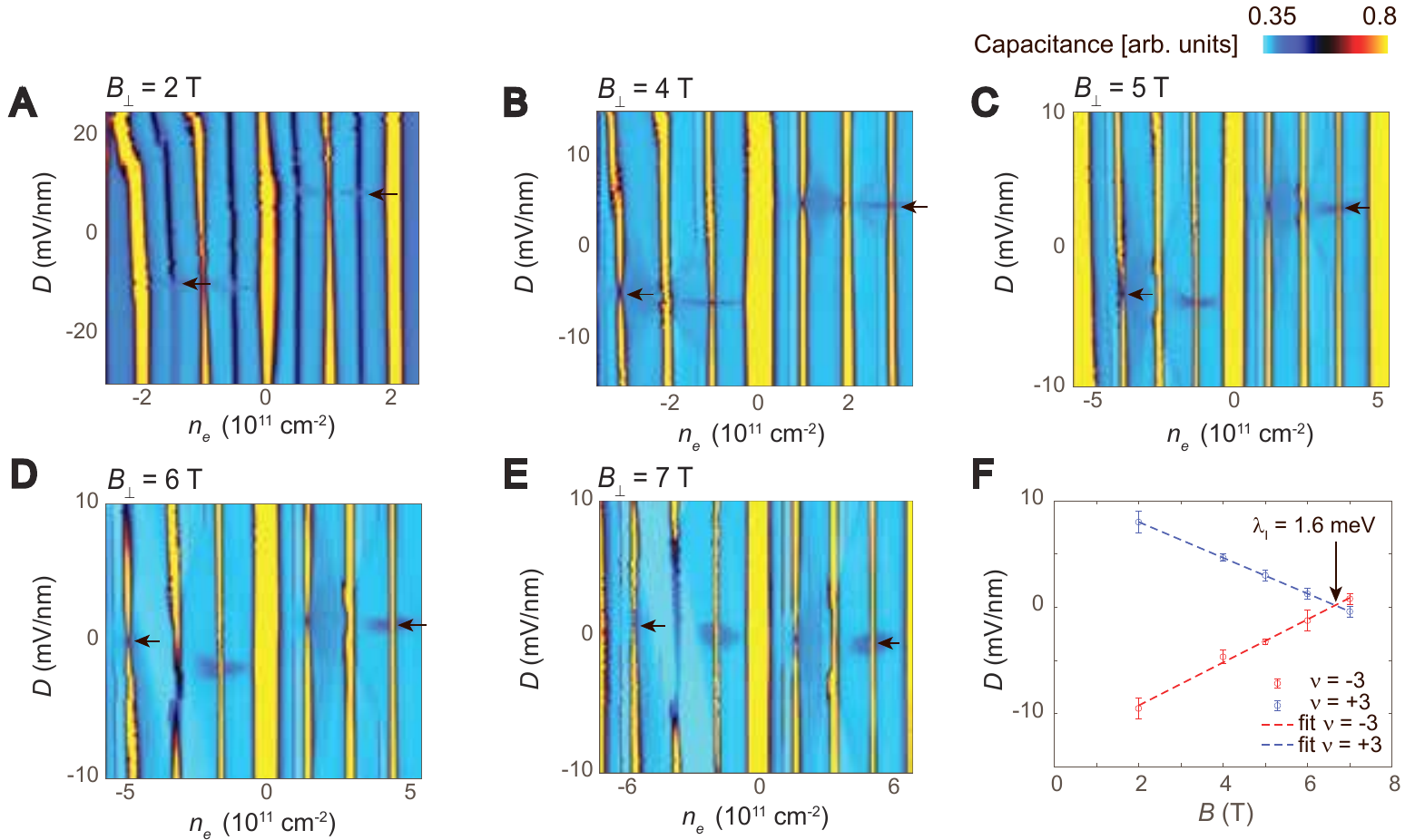}
    \caption{\textbf{Measurement of Ising SOC from transitions in the lowest Landau level (LL):}
    \textbf{(A - E)} $D$ - $n_e$ phase diagrams for different out-of-plane magnetic fields $B_\perp$ at low displacement fields. 
    The black arrows mark the orbital transitions of the $\nu$ = $\pm$ 3 LL used to determine the Ising SOC. 
    \textbf{(F)} $B_\perp$ dependent transitions plotted against $D$. The dashed lines are fits to the data. 
    The Ising SOC is calculated from the crossing point of the $\nu$ = $\pm$ 3 lines\cite{Island2019}. 
    Data from this device taken at higher temperatures and without in-plane magnetic fields was previously described in Ref. \cite{Zhang2023}.}
    \label{fig:S1}. 
\end{figure*}

\begin{figure*}[h]
    \centering
    \includegraphics[width = 150mm]{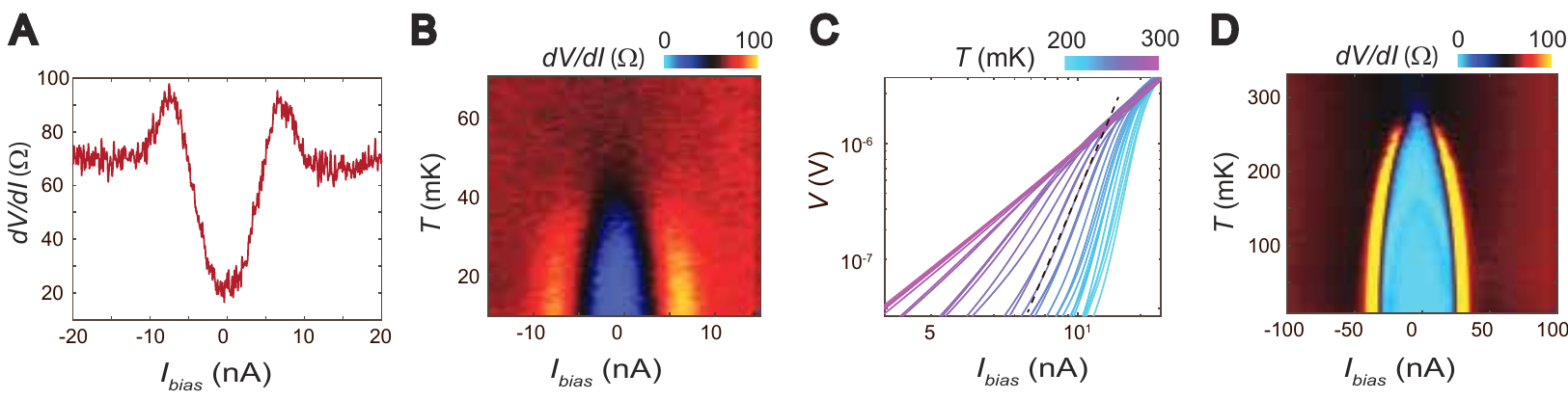}
    \caption{\textbf{Additional bias and temperature dependence of SC$_1$ and SC$_2$.}
    \textbf{(A)} linecut of the $dV/dI$ spectrum of Fig.~\ref{fig:1}F at zero field and $n_e$ = -8.7 $\cdot$ 10$^{11}$ cm$^{-2}$ and $D$ = 0.95 V/nm.
    \textbf{(B)} Temperature dependence of panel A. 
    \textbf{(C)} I-V characteristics extracted from temperature dependent non-linear transport data of SC$_2$ in \textbf{(D)} at $n_e$ = -7.3 $\cdot$ 10$^{11}$ cm$^{-2}$ and $D$ = 1.15 V/nm. 
    The dashed line in C is a fit where $V$ $\sim$ $I^3$ which we define as the BKT-transition temperature.}
    \label{fig:S2}
\end{figure*}

\begin{figure*}
    \centering
    \includegraphics[width = 150mm]{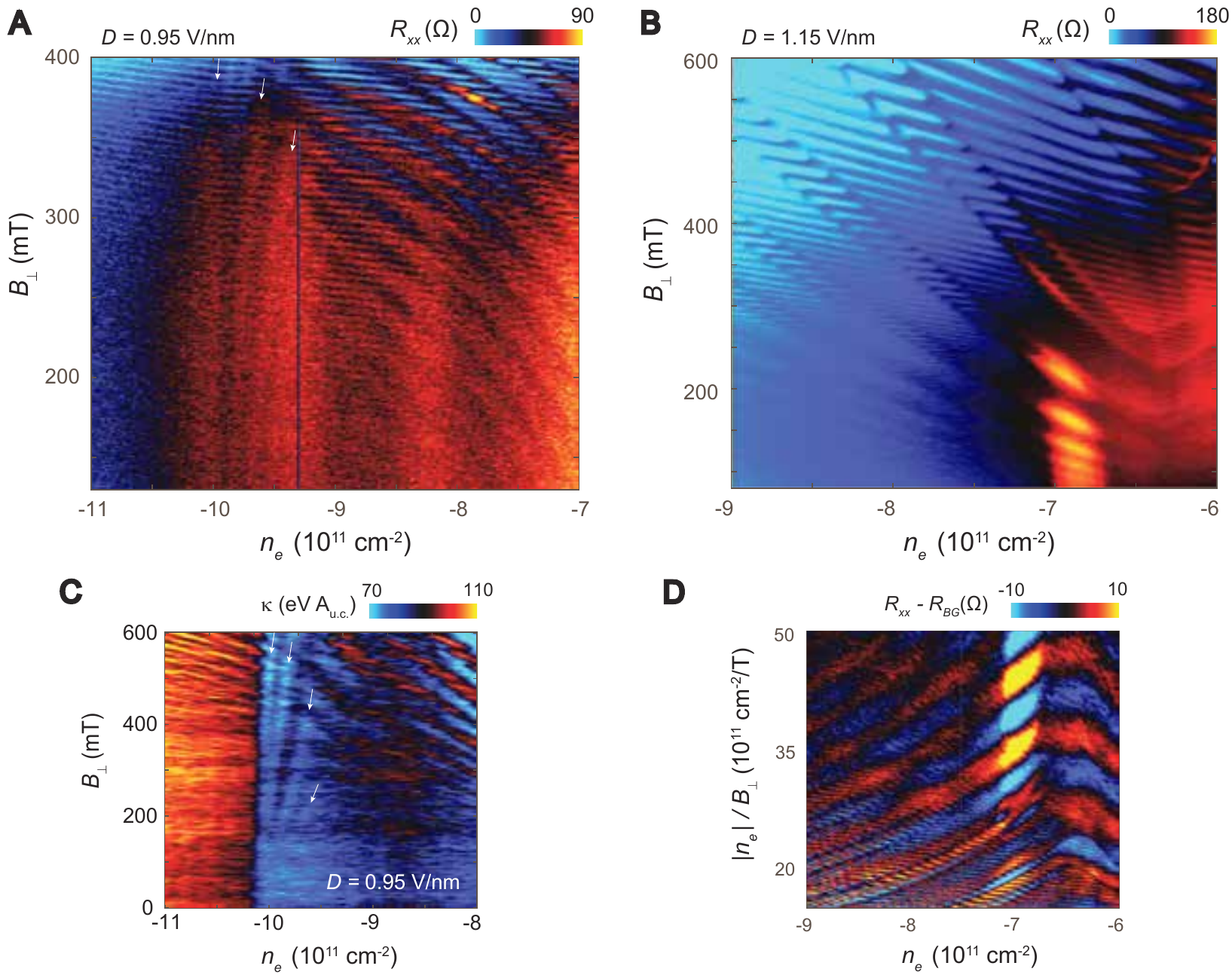}
    \caption{\textbf{Raw quantum oscillation data.}
    \textbf{(A - B)} Shubnikov-de-Haas data taken at $D$ = 0.95 V/nm and 1.15 V/nm, respectively. 
    \textbf{(C)} $B_\perp$ dependent inverse compressibility determined from penetration field capacitance data at $D$ = 0.95 V/nm. The white arrows  in A and C indicate the Landau fan associated with the small electron annulus in the Ising$_{2,2,-2}$ phase.
    \textbf{(D)} Same data as in panel B plotted as density $n_e$ over $B_\perp$. 
    An average background is subtracted from the raw data to visually enhance the quantum oscillations and illustrate no obvious change of quantum oscillations across the COS around $n_e$ = -6.9 $\cdot$ 10$^{11}$cm$^{-2}$.}
    \label{fig:S3}
\end{figure*}

\begin{figure*}
    \centering
    \includegraphics[width = 150mm]{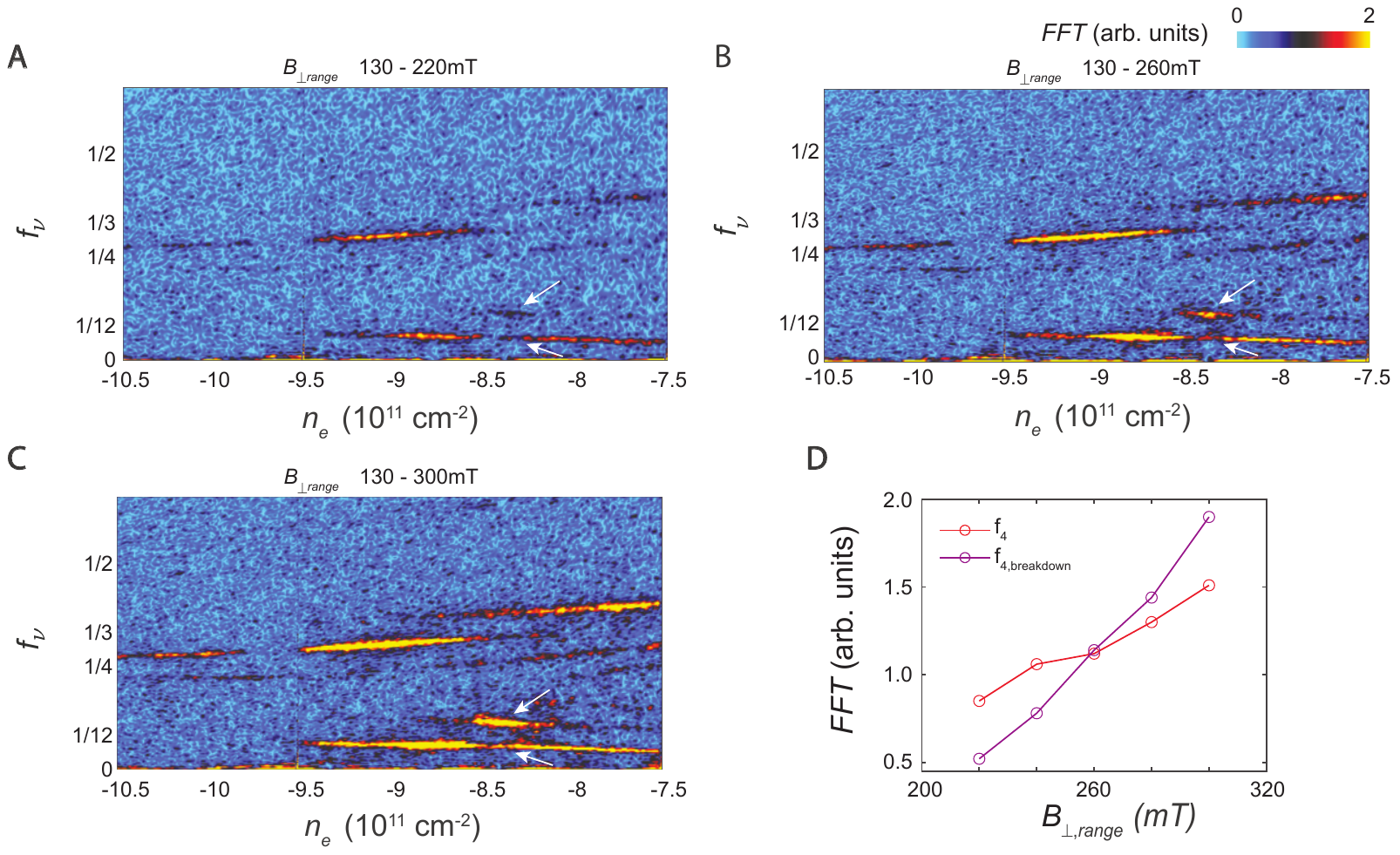}
    \caption{\textbf{Magnetic breakdown between the small Fermi pockets.}
    \textbf{(A - C)} Fourier transformation of the data in Fig.~\ref{fig:S3}A taken over different field ranges. 
    The white arrows mark $f_4$ and $f_{4,breakdown}$. 
    \textbf{(D)} Field range dependent FFT amplitude averaged over densities between -8.49 $\cdot$ 10$^{11}$ cm$^{-2}$ to -8.3 $\cdot$ 10$^{11}$ cm$^{-2}$. 
    The relative weight between $f_4$ and $f_{4,breakdown}$ increases in favor of the latter for a larger and higher $B_{\perp,range}$, illustrating that $f_{4,breakdown}$ is due to magnetic breakdown between the fermi pockets associated with the frequency $f_4$.}
    \label{fig:S4}
\end{figure*}

\begin{figure*}
    \centering
    \includegraphics[width = 150mm]{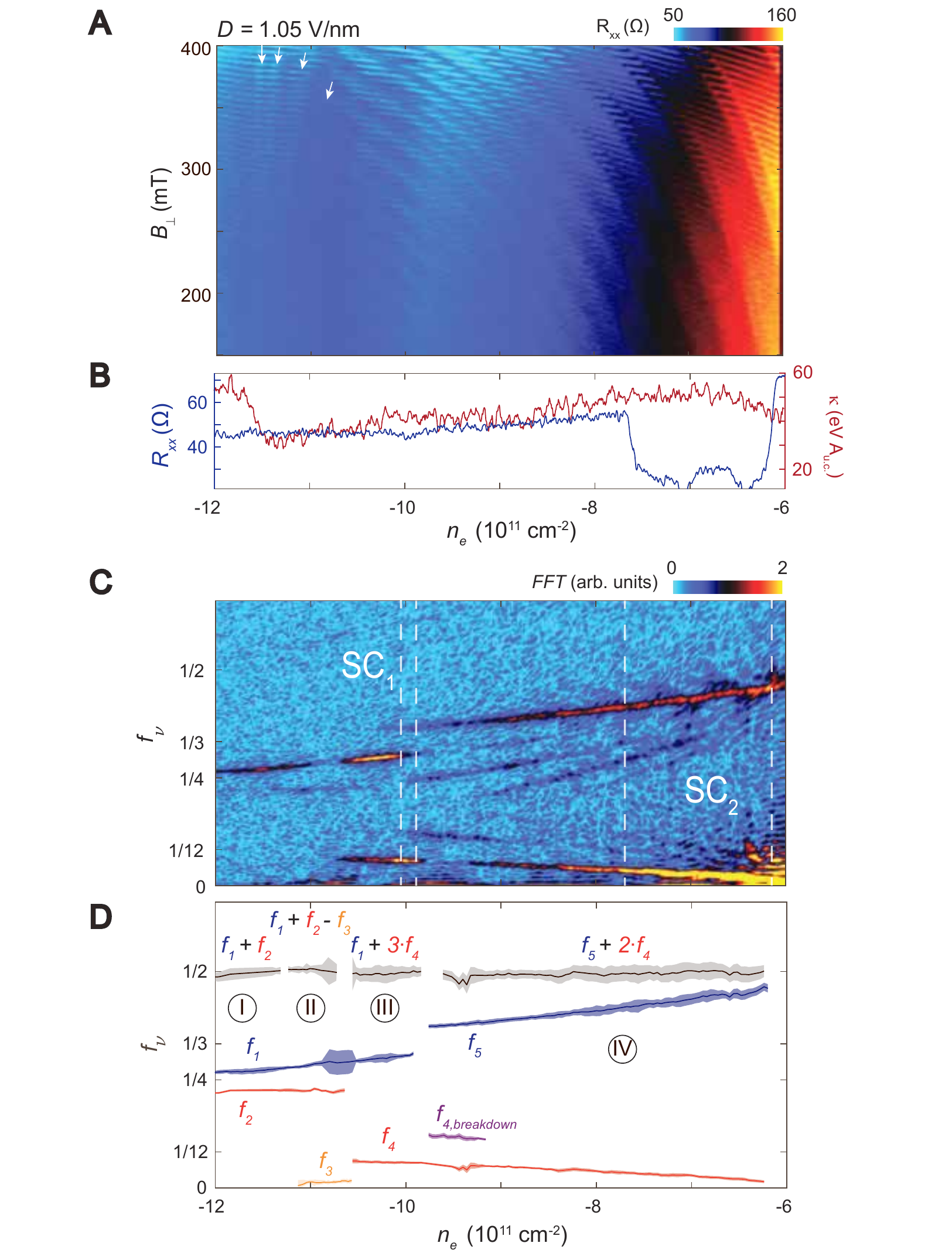}
    \caption{\textbf{Shubnikov-de-Haas measurements and fermiology analysis at $D$ = 1.05 V/nm.}
    \textbf{(A)} raw quantum oscillation data which is used to compute the Fourier transform in panel C. 
    The white arrows indicate electron like quantum oscillations associated with $f_3$. 
    \textbf{(B)} comparison of $R_{xx}$ and inverse compressibility $\kappa$ at zero magnetic field. 
    \textbf{(C)} Fourier transform of the data in A for fields from 150 - 300 mT similar to main text Fig.~\ref{fig:2}B. 
    The region of SC$_{1,2}$ are indicated by the white dashed lines. As SC$_1$ is not well formed and is barely visible in $R_{xx}$, one should comfirm its location with the 2D resistance map in Fig.~\ref{fig:S12}. 
    Note: due to the higher field range used here compared to the main text, additional peaks associated with magnetic breakdown effects between different pockets of $f_{4,5}$ are apparent. 
    \textbf{(D)} Schematic of panel C with main frequencies extracted via Gaussian fits to the data ($f_3$ is not well captured by fits and extracted manually here). 
    The same phases as in the main text Fig.~\ref{fig:2} are observed, most importantly showing the transition to the nematic N$_{2,4}$ phase.}
    \label{fig:S5}
\end{figure*}

\begin{figure*}
    \centering
    \includegraphics[width = 150mm]{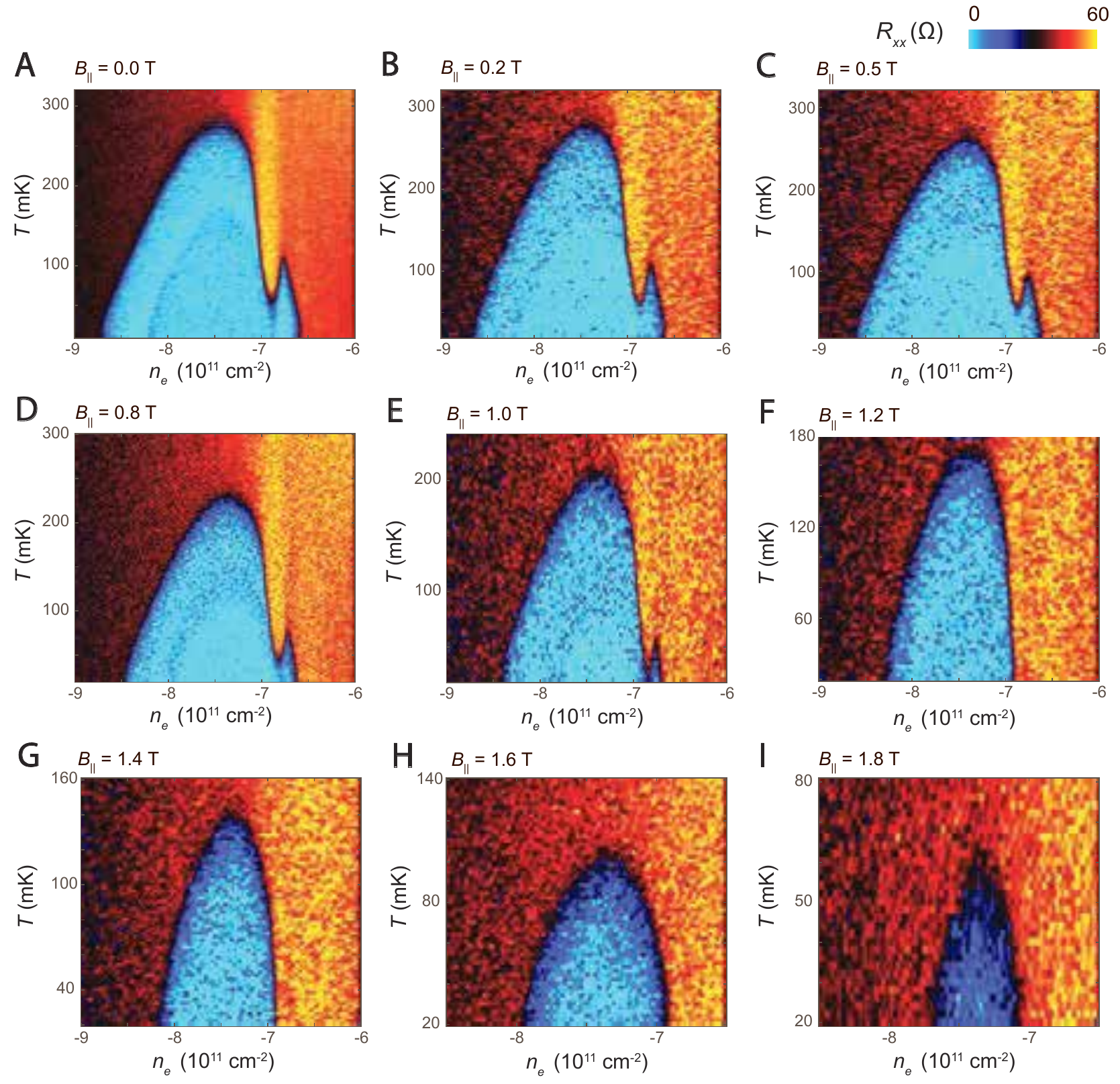}
    \caption{\textbf{In-plane field dependence of SC$_2$ at D = 1.15 V/nm.}
    \textbf{(A - I)} $T_c$ domes for different in-plane magnetic fields $B_{\parallel}$ up to 1.8 T.}
    \label{fig:S6}
\end{figure*}

\begin{figure*}
    \centering
    \includegraphics[width = 150mm]{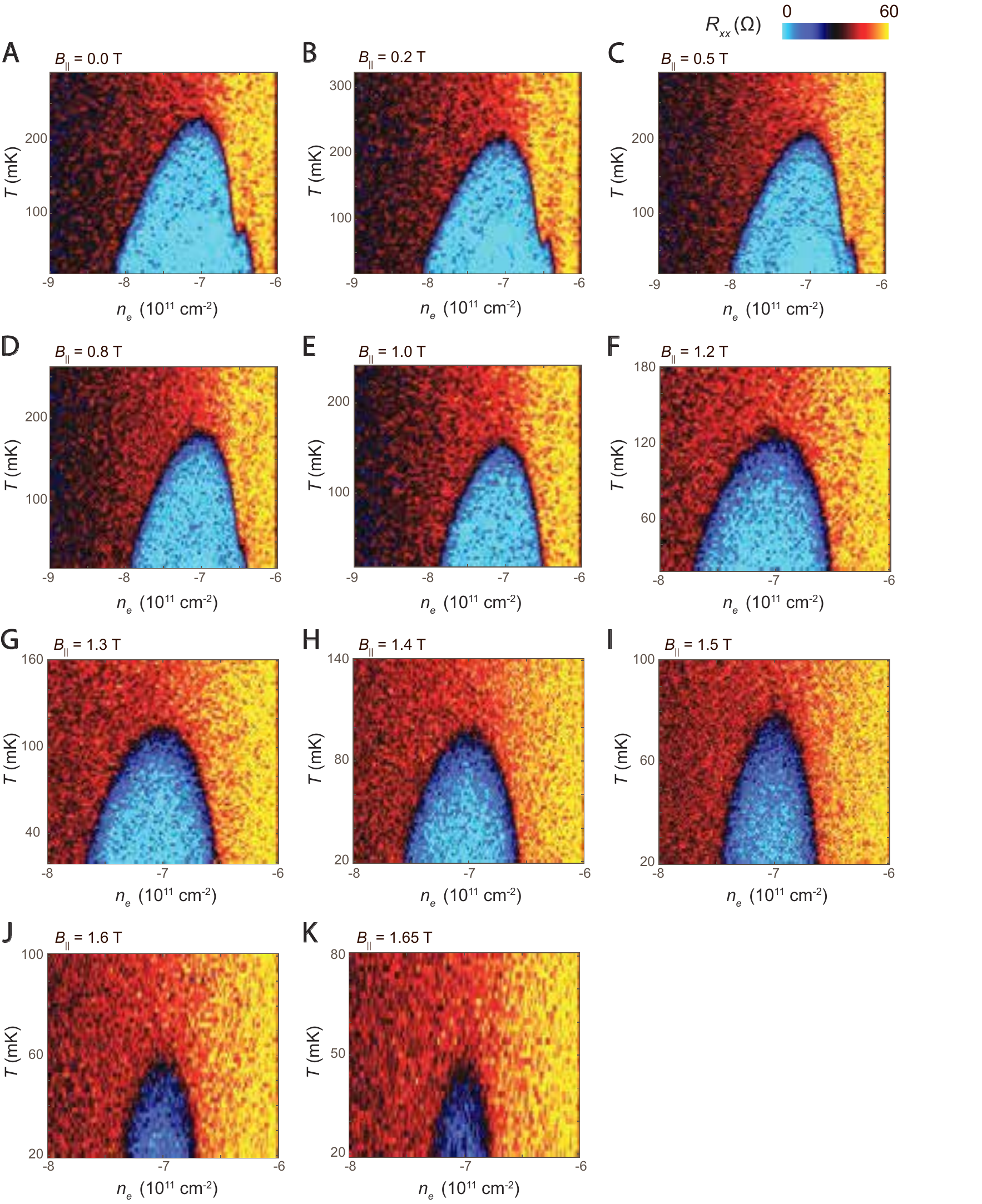}
    \caption{\textbf{In-plane field dependence of SC$_2$ at D = 1.1 V/nm.}
    \textbf{(A - K)} $T_c$ domes for different in-plane magnetic fields $B_{\parallel}$ up to 1.65 T.}
    \label{fig:S7}
\end{figure*}

\begin{figure*}
    \centering
    \includegraphics[width = 150mm]{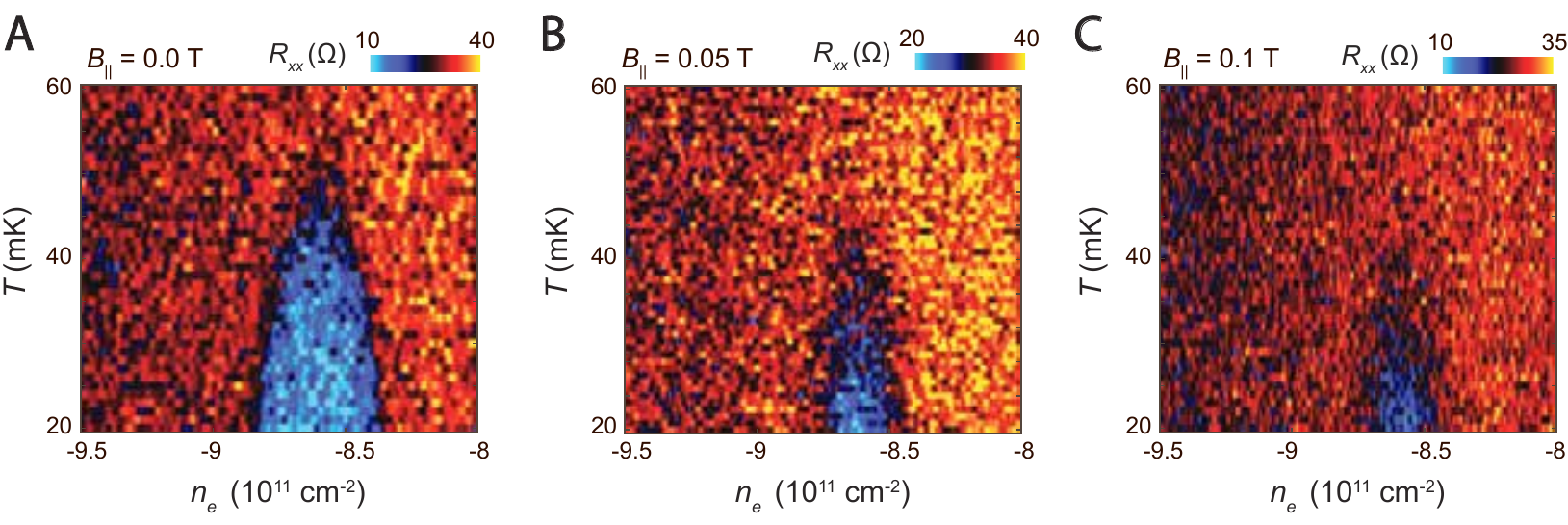}
    \caption{\textbf{In-plane field dependence of SC$_1$ at D = 0.94 V/nm.} 
    \textbf{(A - C)} $T_c$ domes for different in-plane magnetic fields $B_{\parallel}$ up to 0.1 T.}
    \label{fig:S8}
\end{figure*}

\begin{figure*}
    \centering
    \includegraphics[width = 150mm]{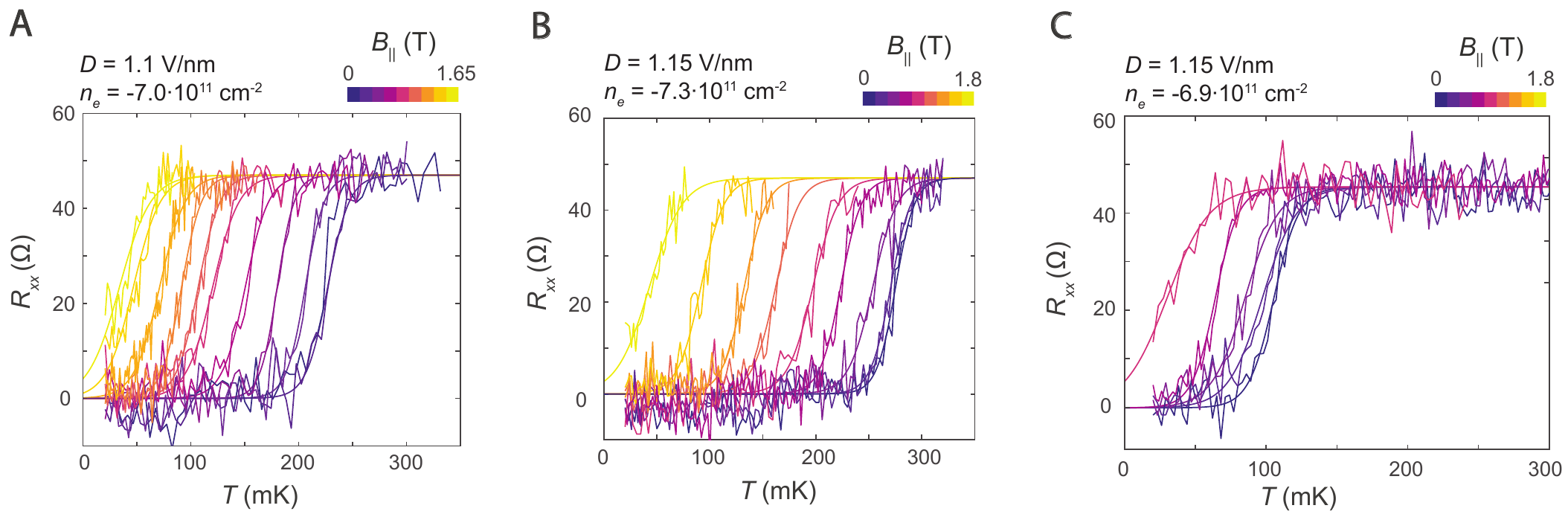}
    \caption{\textbf{Critical temperature extraction from superconducting domes.} R-T linecuts of the data of fig. \ref{fig:S6}, \ref{fig:S7}, \ref{fig:S8} from which in-plane field dependent critical temperatures are extracted. Both raw data and fits to the data are plotted. \textbf{A-C} display the three different $n_e$, $D$ values. Experimental values in fig. \ref{fig:3} B are taken from panel B.}
    \label{fig:S17}
\end{figure*}

\begin{figure*}
    \centering
    \includegraphics[width = 150mm]{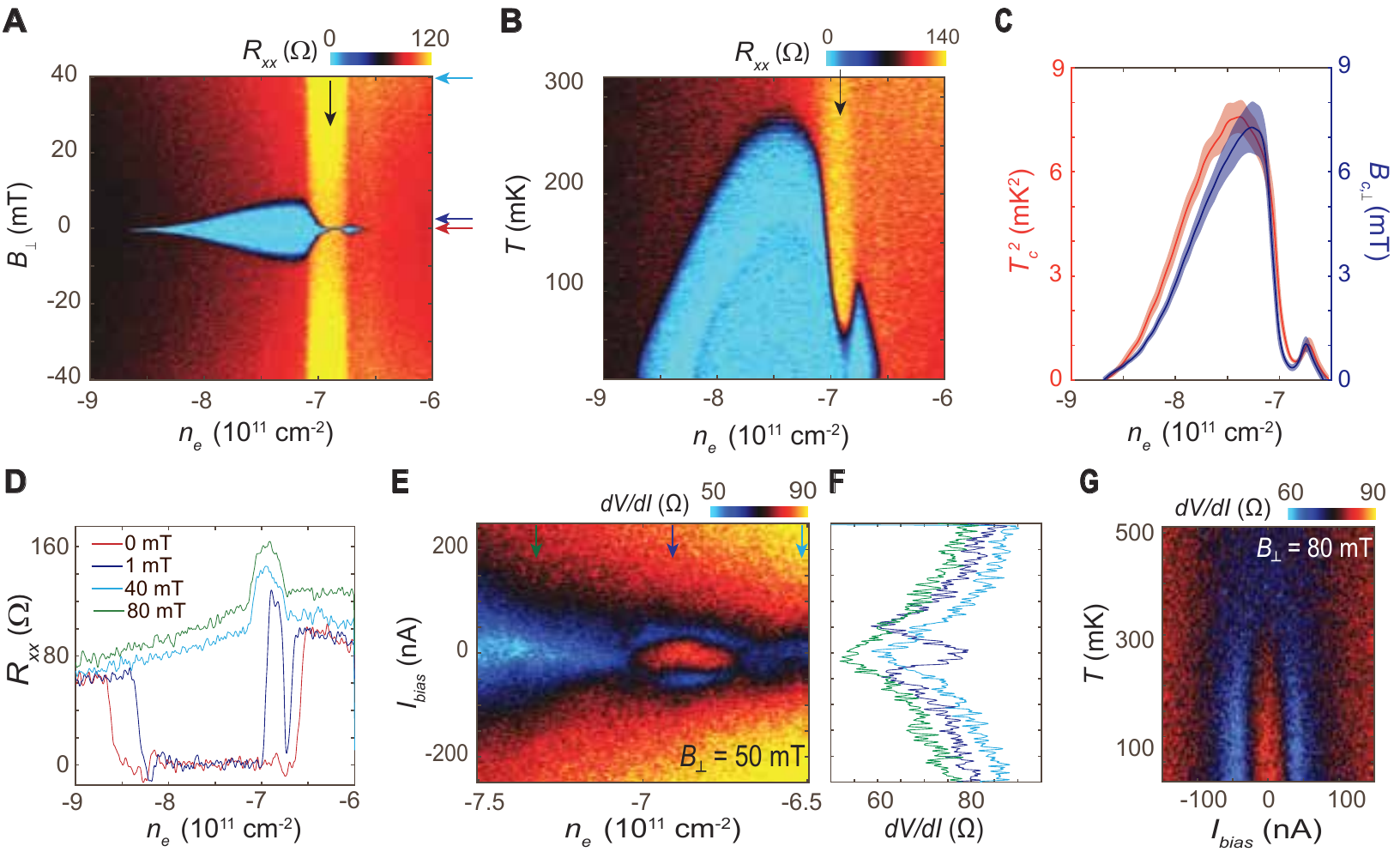}
    \caption{\textbf{Competing phase in the superconducting regime.} 
    \textbf{(A)} $n_e$ and $B_\perp$ dependence of SC$_2$ at $D$ = 1.15 V/nm. 
    \textbf{(B)} $n_e$ and temperature dependence of SC$_2$ at the same displacement field.
    \textbf{(C)} Comparison of $T_c^2$ and $B_{c,\perp}$ extracted from fits to the data in A, B. 
    Shaded regions represent error bars representing the width of the SC transition. 
    \textbf{(D)} $n_e$ linecuts of A for different $B_\perp$. 
    \textbf{(E)} DC current bias dependence of COS. 
    \textbf{(F)} Linecuts of E at the $n_e$ values marked by the arrows in corresponding color. 
    \textbf{(G)} Temperature - current bias dependence at $n_e$ = -6.95 $\cdot$ 10$^{11}$ cm$^{-2}$, $D$ = 1.15 V/nm and $B_\perp$ = 80 mT.}
    \label{fig:S9}
\end{figure*}

\begin{figure*}
    \centering
    \includegraphics[width = 150mm]{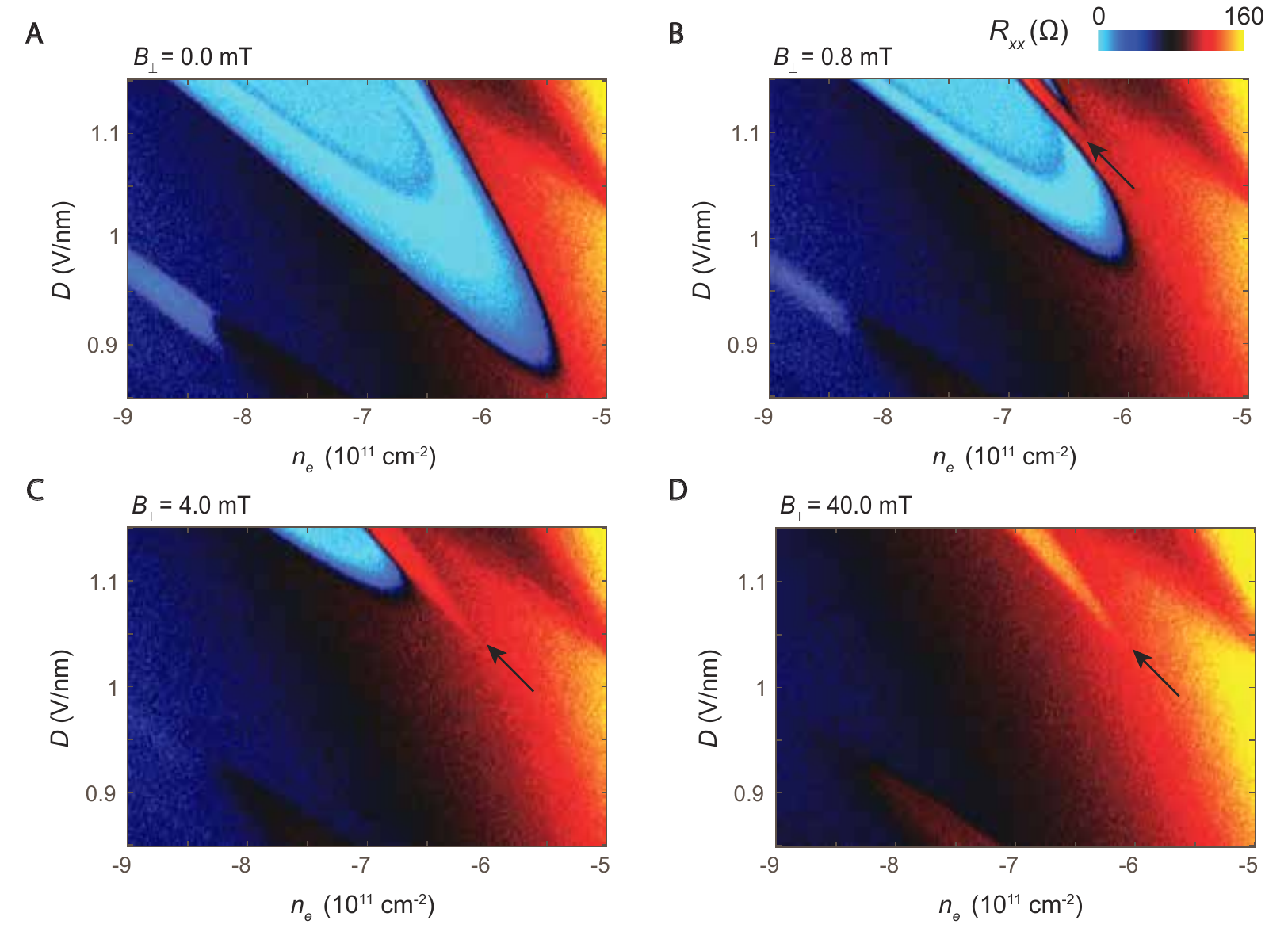}
    \caption{\textbf{COS in the $n_e$ - D phase diagram.}
    \textbf{(A - D)} phase diagrams for vanishing as well as small out of plane magnetic fields. 
    Black arrows indicate the feature assoicated with the COS.}
    \label{fig:S10}
\end{figure*}

\begin{figure*}
    \centering
    \includegraphics[width = 150mm]{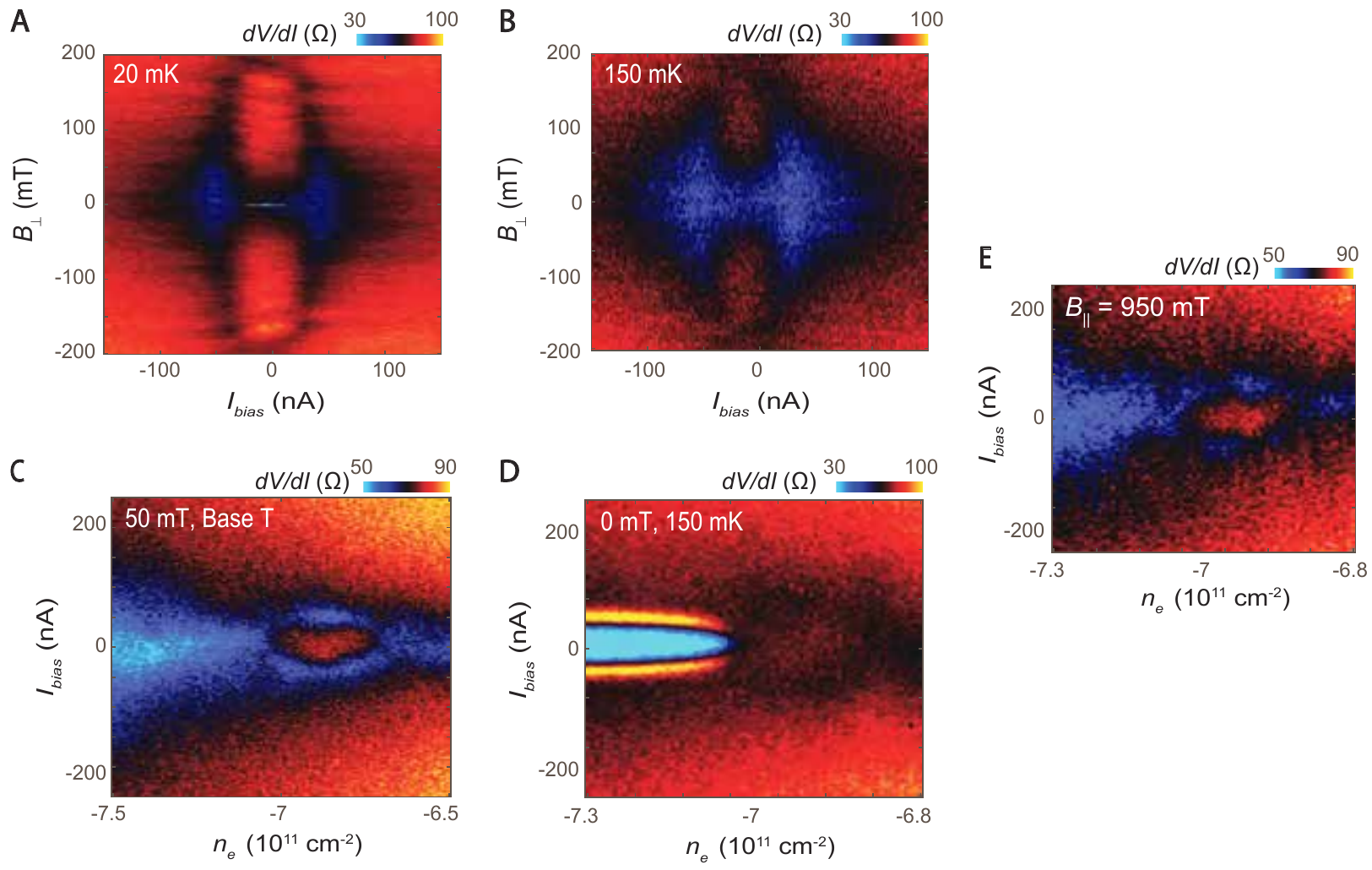}
    \caption{\textbf{Additional data of the COS at $D$ = 1.15 V/nm.}
    \textbf{(A - B)} Out-of plane field dependence of $dV/dI$ and $n_e$ $\approx$ -6.9 $\cdot$ 10$^{11}$ cm$^{-2}$ at base temperature and 150 mK. 
    \textbf{(C - D)} Comparison of $dV/dI$ at base temperature, B$_\perp$ = 50 mT with $T$ = 150 mK and B$_\perp$ = 0 mT. 
    \textbf{E} $dV/dI$ at high in-plane field and  B$_\perp$ = 50 mT showing that COS is only weakly dependent on $B_{\parallel}$.}
    \label{fig:S11}
\end{figure*}

\begin{figure*}
    \centering
    \includegraphics[width = 150mm]{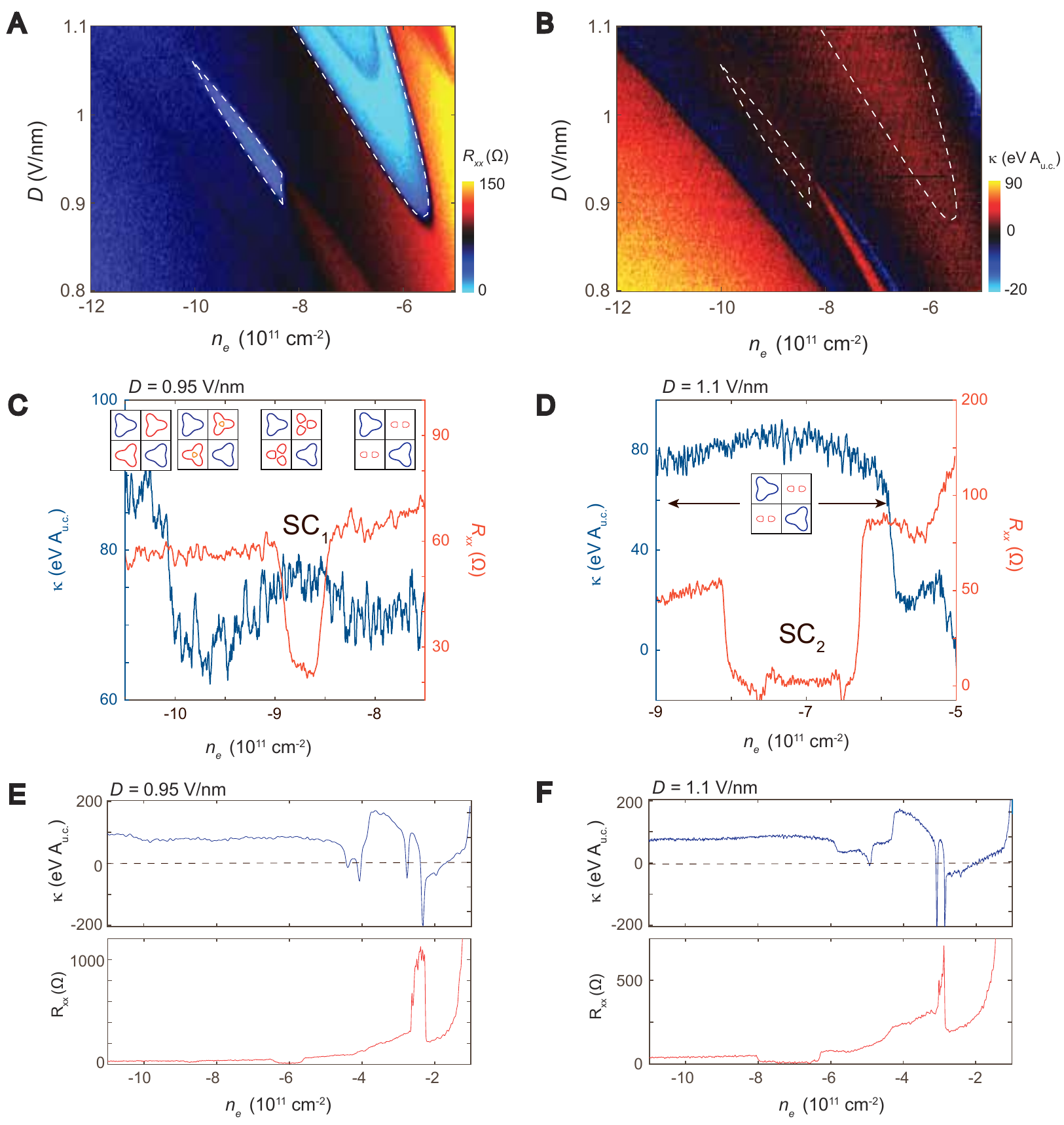}
    \caption{\textbf{Comparison of transport and penetration field capacitance.}
    \textbf{(A - B)} $n_e$-$D$ phase diagram of $R_{xx}$ and $\kappa$ over a larger density range than Fig.~\ref{fig:4}D, including SC$_1$. 
    The superconducting regions are overlaid from transport onto the inverse compressibilit map via the white dashed line. 
    \textbf{(C)} linecuts of $R_{xx}$ and $\kappa$ at $D$ = 0.95 V/nm across SC$_1$. 
    The noise of $\kappa$ gives an upper bound of $\sim$ 300 mK on a possible chemical potential jump due to a first order phase transition below our experimental resolution. The fermiology is added as determined in Fig.~\ref{fig:2}. 
    \textbf{(D)} same as Fig.~\ref{fig:4}C for a different $D$ = 1.1 V/nm. 
    \textbf{(E - F)} $\kappa$ (top panel) and $R_{xx}$ (bottom panel) linecuts at $D$ = 0.95 V/nm and 1.1 V/nm. 
    At low $|n_e|$, several negative spikes in $\kappa$ become apparent---illustrating the sensitivity of $\kappa$ to first order phase transitions. The dashed lines marks zero as guide to the eye.}
    \label{fig:S12}
\end{figure*}


\begin{figure*}
    \centering
    \includegraphics[width = 150mm]{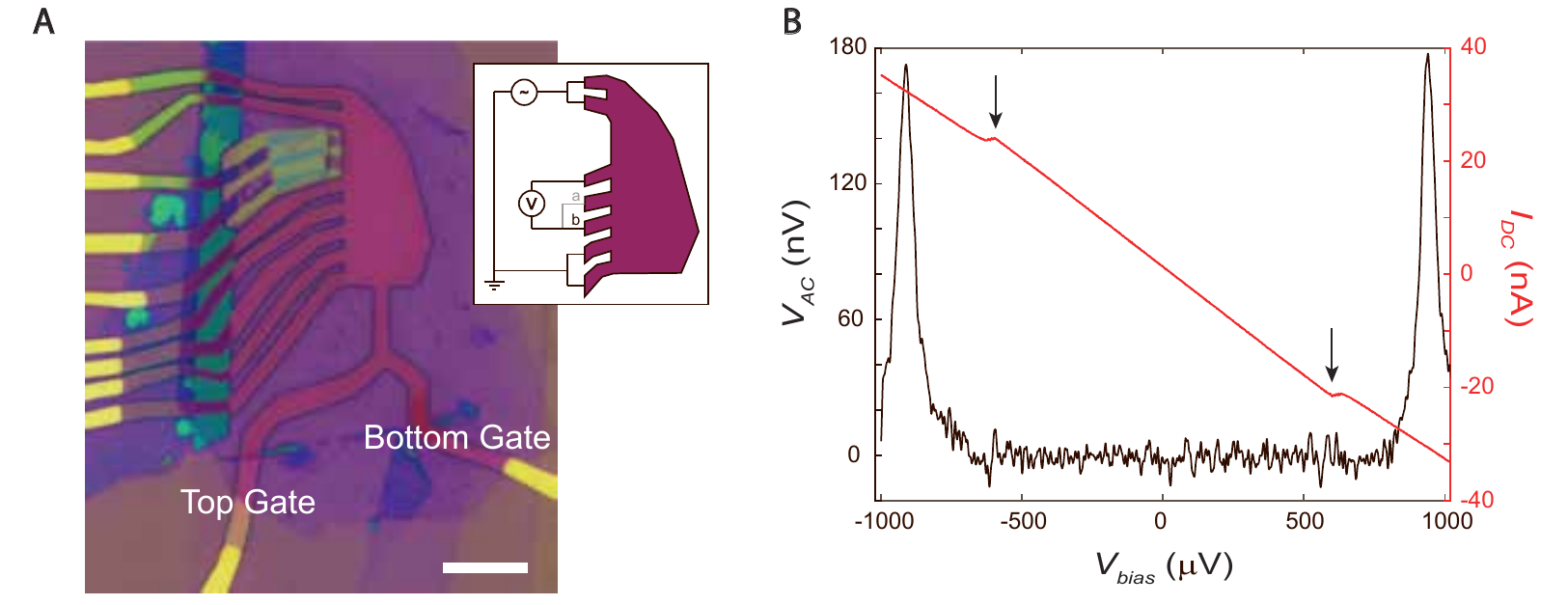}
    \caption{\textbf{Sample image and raw data of the current-voltage characteristics of the superconducting state.}
    \textbf{(A)} image of the sample. 
    The scale bar is 10 $\mu$m. 
    The inset shows the measurement setup used for transport measurements (contacts a and b are both used as negative voltage contact in different sets of measurements). 
    \textbf{(B)} Voltage bias dependence of the raw voltage and DC current data at $D$ = 1.15 V/nm, $n_e$ = -7.3 $\cdot$ 10$^{11}$ cm$^{-2}$, deep within SC$_2$. 
    An AC-excitation of 1 nA is used for to measure the voltage drop $V_{AC}$ across the sample region while ramping the DC voltage bias, $V_{bias}$. 
    The arrows indicate DC current jumps due to the superconducting transition of the dual gated bilayer graphene leads---uncorrelated with the voltage drop across the bulk of the sample.}
    \label{fig:S13}
\end{figure*}

\begin{figure*}
    \centering
    \includegraphics[width = 150mm]{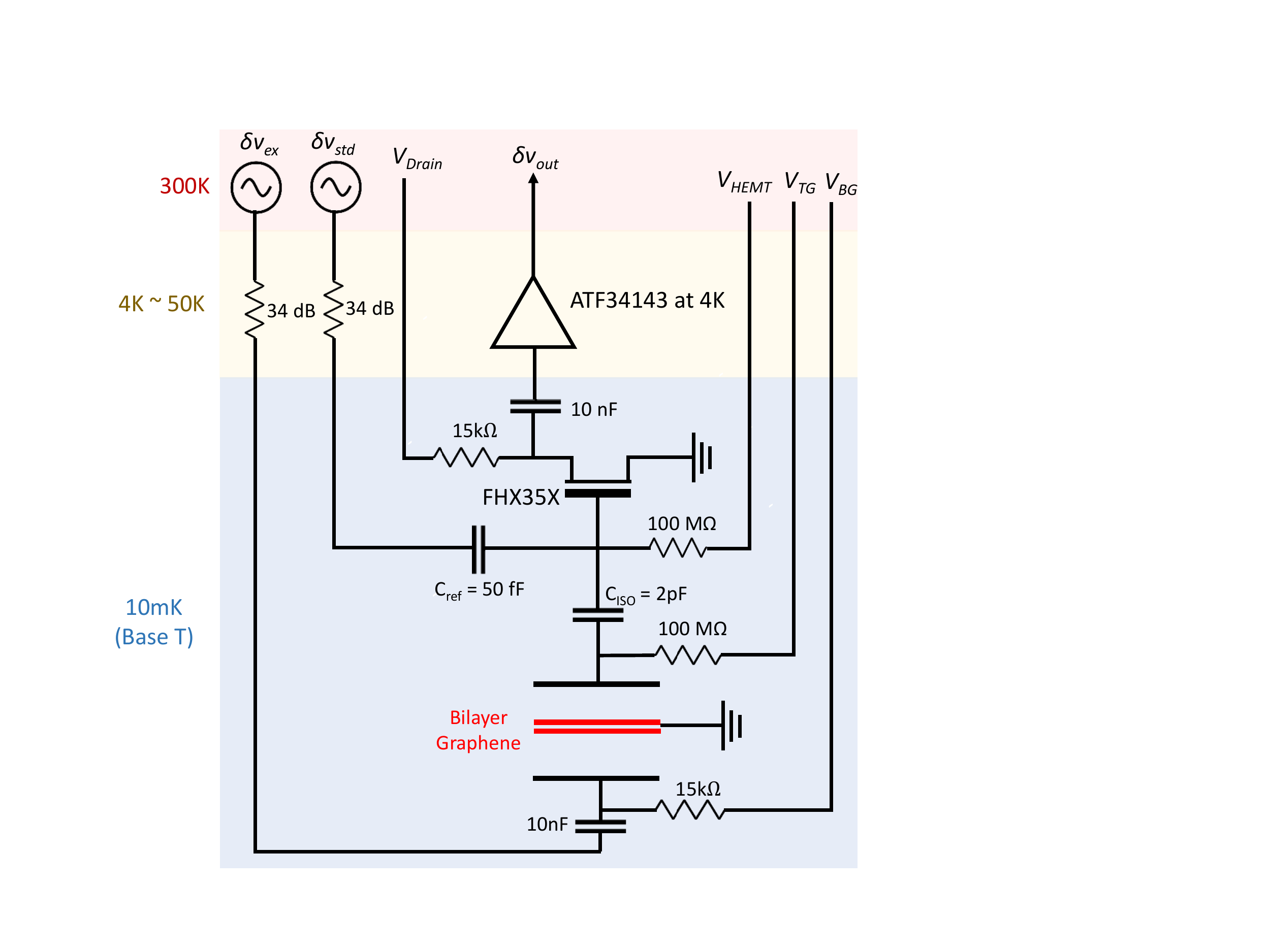}
    \caption{\textbf{Circuit schematic of the penetration field capacitance measurements.}}
    \label{fig:S14}
\end{figure*}

\begin{figure*}
    \centering
    \includegraphics[width = 150mm]{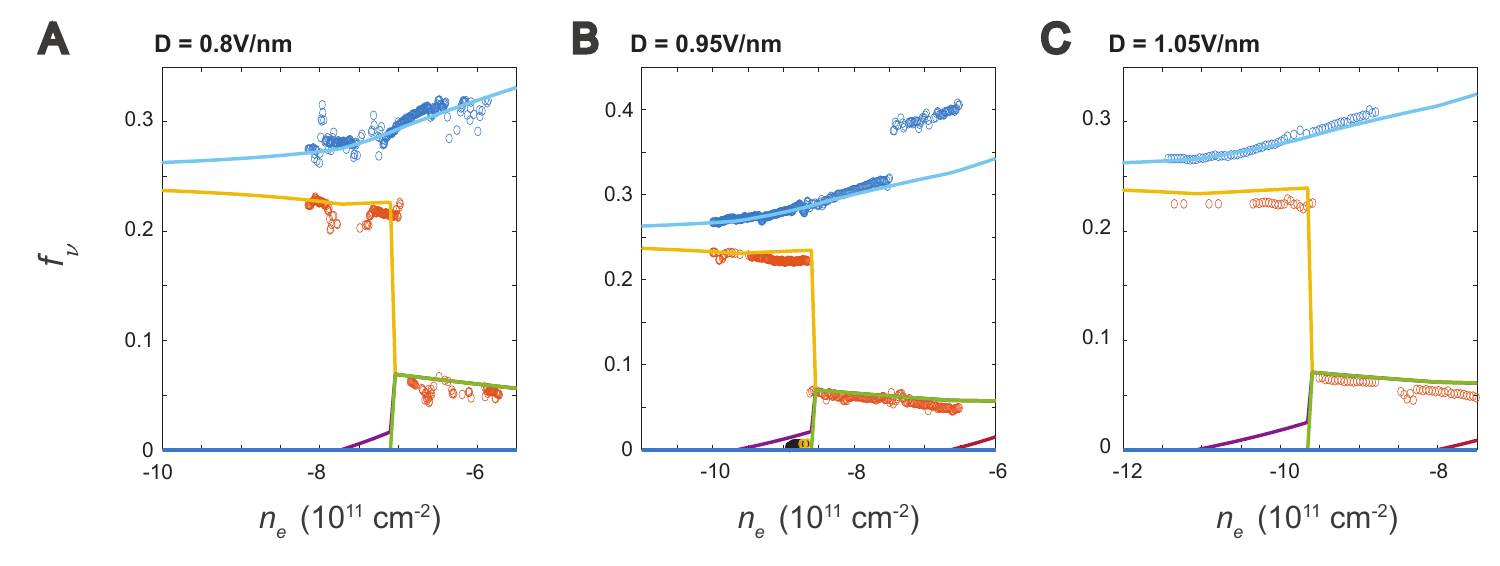}
    \caption{\textbf{Bandstructure fits to quantum oscillations.} We find $\gamma_0$ = 2880 meV, $\gamma_1$ = 361 meV, $\gamma_3$ = 323 meV, $\gamma_4$ = 30 meV, $\delta$ = 13 meV with Ising SOC set to $\lambda_I$ = 1.6 meV from fits to single particle calculations. These values are used to calculate the in-plane orbital pair breaking.}
    \label{fig:S19}
\end{figure*}

\begin{figure*}
    \centering
    \includegraphics[width = 75mm]{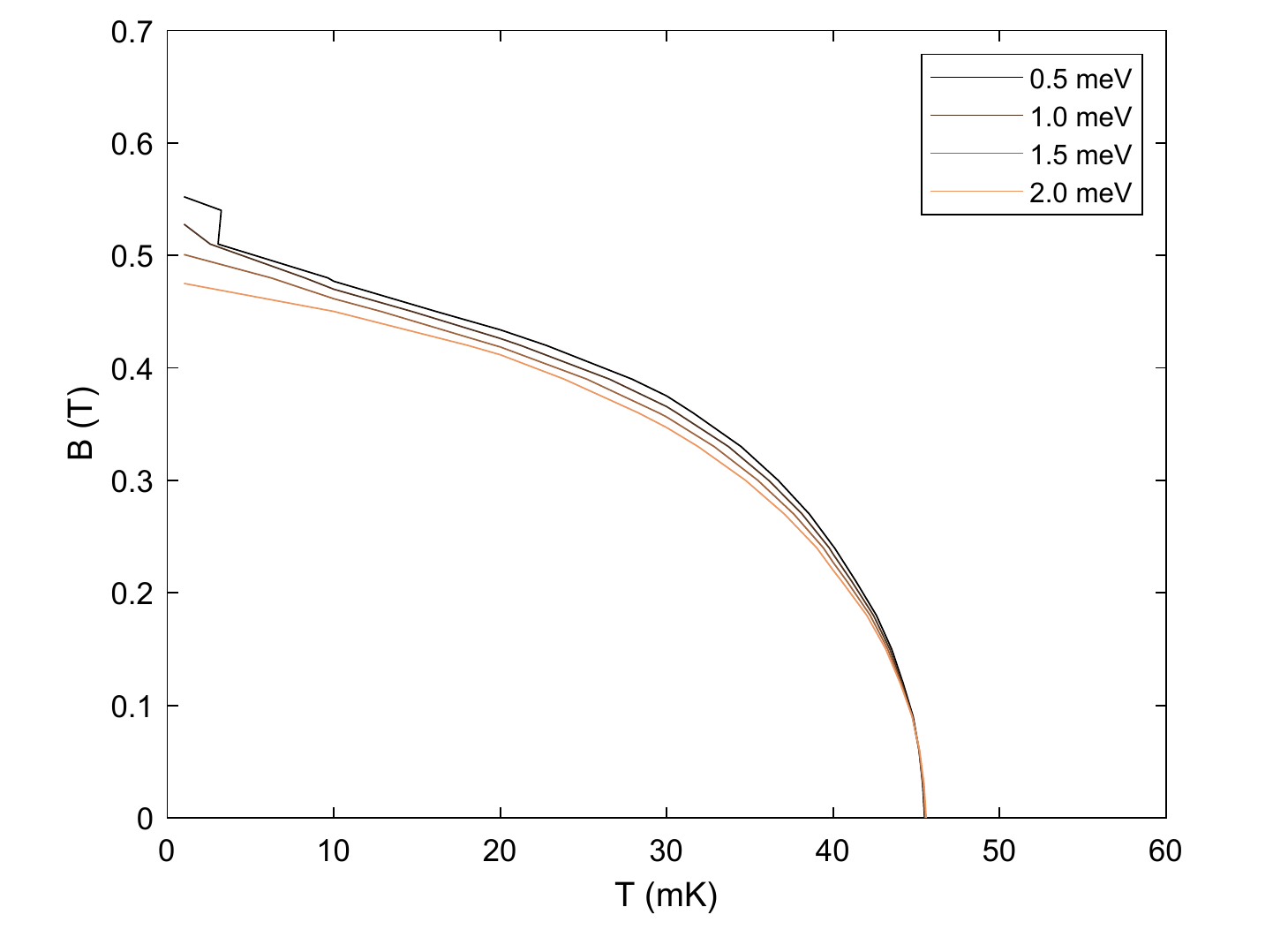}
    \caption{\textbf{In-plane critical fields including orbital and Rashba SOC effects.} 
 Analogous calculations to those of Fig. \ref{fig:3}c of the main text, including both orbital depairing and the effect of Rashba spin orbit coupling with values ranging from from 0.5 - 2 meV.}
    \label{fig:S20}
\end{figure*}

\end{document}